\documentclass[twocolumn]{aa}
\usepackage{txfonts}
\usepackage{multirow}
\usepackage[linktocpage=true,colorlinks=true,linkcolor=blue,citecolor=blue,urlcolor=blue]{hyperref}

\usepackage[switch]{lineno}
\usepackage{amsmath}
\usepackage{amssymb}
\usepackage{natbib}
\usepackage{graphicx}

\newcommand*\patchAmsMathEnvironmentForLineno[1]{
  \expandafter\let\csname old#1\expandafter\endcsname\csname #1\endcsname
  \expandafter\let\csname oldend#1\expandafter\endcsname\csname end#1\endcsname
  \renewenvironment{#1}
     {\linenomath\csname old#1\endcsname}
     {\csname oldend#1\endcsname\endlinenomath}}
\newcommand*\patchBothAmsMathEnvironmentsForLineno[1]{
  \patchAmsMathEnvironmentForLineno{#1}
  \patchAmsMathEnvironmentForLineno{#1*}}
\AtBeginDocument{
\patchBothAmsMathEnvironmentsForLineno{equation}
\patchBothAmsMathEnvironmentsForLineno{align}
\patchBothAmsMathEnvironmentsForLineno{flalign}
\patchBothAmsMathEnvironmentsForLineno{alignat}
\patchBothAmsMathEnvironmentsForLineno{gather}
\patchBothAmsMathEnvironmentsForLineno{multline}
}

\newcommand{\om}{\Omega_{\rm M}}

\usepackage{color}

\begin{document}
\title{Extending the supernova Hubble diagram to $z \sim 1.5$ with the Euclid space mission}
\titlerunning{Distances to high redshift supernovae with Euclid}
\authorrunning{P.~Astier et al.}

\author{
P.~Astier\inst{1}
\and
C.~Balland\inst{1}
\and
M.~Brescia\inst{2}
\and
E.~Cappellaro\inst{3}
\and
R.~G.~Carlberg\inst{4}
\and
S.~Cavuoti\inst{5}
\and
M.~Della~Valle\inst{2,6}
\and
E.~Gangler\inst{7}
\and
A.~Goobar\inst{8}
\and
J.~Guy\inst{1}
\and
D.~Hardin\inst{1}
\and
I.~M.~Hook\inst{9,10}
\and
R.~Kessler\inst{11,12}
\and
A.~Kim\inst{13}
\and
E.~Linder\inst{14}
\and
G.~Longo\inst{5}
\and
K.~Maguire\inst{9,15}
\and
F.~Mannucci\inst{16}
\and
S.~Mattila\inst{17}
\and
R.~Nichol\inst{18}
\and
R.~Pain\inst{1}
\and
N.~Regnault\inst{1}
\and
S.~Spiro\inst{9}
\and
M.~Sullivan\inst{19}
\and
C.~Tao\inst{20,21}
\and
M.~Turatto\inst{3}
\and
X.~F.~Wang\inst{21}
\and
W.~M.~Wood-Vasey\inst{22}
}
\institute{
LPNHE, CNRS/IN2P3, Universit\'e Pierre et Marie Curie Paris 6, Universit\'e Denis Diderot Paris 7, 4 place Jussieu, 75252 Paris Cedex 05, France
\and
INAF, Capodimonte Astronomical Observatory, via Moiariello 16, I-80131 Naples, Italy
\and
INAF Osservatorio Astronomico di Padova, Vicolo dell'Osservatorio 5, 35122 Padova, Italy
\and
Department of Astronomy and Astrophysics, University of Toronto, 50 St. George Street, Toronto ON M5S 3H4, Canada
\and
Dept. of Physics, University Federico II, via Cinthia I-80126 Naples, Italy
\and
International Center for Relativistic Astrophysics, Piazza Repubblica, 10, I-65122, Pescara, Italy
\and
Clermont Université, Université Blaise Pascal, CNRS/IN2P3, Laboratoire de Physique Corpusculaire, BP 10448, F-63000 Clermont-Ferrand, France
\and
Albanova University Center, Department of Physics, Stockholm University, Roslagstullsbacken 21, 106 91 Stockholm, Sweden
\and
Department of Physics (Astrophysics), University of Oxford, DWB, Keble Road, Oxford OX1 3RH, UK
\and
INAF - Osservatorio Astronomico di Roma, via Frascati 33, 00040 Monteporzio (RM), Italy
\and
Department of Astronomy and Astrophysics, University of Chicago, 5640 South Ellis Avenue, Chicago, IL 60637
\and
Kavli Institute for Cosmological Physics, University of Chicago, 5640 South Ellis Avenue Chicago, IL 60637
\and
LBNL, 1 Cyclotron Rd, Berkeley, CA 94720, USA
\and
University of California, Berkeley, CA 94720 USA
\and
European Southern Observatory, Karl-Schwarzschild-Str. 2, 85748 Garching bei München, Germany
\and
INAF, Osservatorio Astrofisico di Arcetri, Largo E.Fermi 5, I-50125 Firenze, Italy
\and
Finnish Centre for Astronomy with ESO (FINCA), University of Turku, V\"ais\"al\"antie 20, FI-21500 Piikki\"o, Finland
\and
Institute of Cosmology \& Gravitation, University of Portsmouth, Portsmouth PO1 3FX, UK
\and
School of Physics and Astronomy, University of Southampton, Southampton, SO17 1BJ, UK
\and
CPPM, Universit\'e Aix-Marseille, CNRS/IN2P3, Case 907, 13288 Marseille Cedex 9, France
\and
Tsinghua center for astrophysics, Physics department, Tsinghua University, 100084 Beijing, China
\and
PITT PACC, Department of Physics and Astronomy, University of Pittsburgh, Pittsburgh, PA 15260, USA.
}

\date{Received Mont DD, YYYY; accepted Mont DD, YYYY}

\abstract{We forecast dark energy constraints that could
  be obtained from a new large sample of Type Ia supernovae
  where those at high redshift are acquired with the Euclid space
  mission. We simulate a three-prong SN survey: a $z<0.35$ nearby
  sample (8000 SNe), a $0.2<z<0.95$ intermediate sample
  (8800 SNe), and a $0.75<z<1.55$ high-$z$ sample
  (1700 SNe). The nearby and intermediate surveys are assumed to be
  conducted from the ground, while the high-z is a joint ground- and
  space-based survey. This latter survey, the "Dark Energy Supernova Infra-Red Experiment" (DESIRE), is designed to fit within 6 months of Euclid observing
  time, with a dedicated observing programme.  We simulate the
  SN events as they would be observed in rolling-search mode
  by the various instruments, and derive the quality of expected
  cosmological constraints. We account for known systematic
  uncertainties, in particular calibration uncertainties including
  their contribution through the training of the supernova model used
  to fit the supernovae light curves. Using conservative assumptions
  and a 1-D geometric {\it Planck} prior, we find that the ensemble of
  surveys would yield competitive constraints: a constant
    equation of state parameter can be constrained to $\sigma(w)=0.022$, and a
    Dark Energy Task Force figure of merit of 203 is found for a two-parameter equation of state. Our
  simulations thus indicate that Euclid can bring a significant
  contribution to a purely geometrical cosmology constraint by
  extending a high-quality SN~Ia Hubble diagram to $z\sim 1.5$.
  We also present other science topics enabled by the DESIRE Euclid observations.}

\maketitle

\section{Introduction}
\label{sec:introduction}

Measuring distances to supernovae Ia (SNe~Ia) at $z\sim 0.5$ allowed
two teams \citep{Riess98b,Schmidt98,Perlmutter99} to independently discover that
the expansion of the universe is now accelerating. The cause of this
acceleration at late times is still unknown and has been attributed to
a new component in the universe admixture: dark energy. One can
describe the acceleration at late times through the equation of
state of dark energy $w$ (namely how its density evolves with redshift
and cosmic time), and the current results are compatible with a static
density, i.e. a cosmological constant \citep[e.g.][]{Betoule14}. Measuring precisely this equation
of state constitutes a crucial step towards understanding the nature of dark energy
\citep{DETF06,ESO-ESA}. Since the discovery of acceleration, we
have narrowed the allowed region of parameter space, from SNe\footnote{In this paper, {\it SN} and {\it SNe} mostly refer to Type Ia supernovae rather 
than supernovae in a more general sense.}
\citep[e.g.][]{Riess04, Astier06, Riess07, WoodVasey07, Kessler09, Conley11, Sullivan11,planck2013-p11,Betoule14,Sako14},
and also with other probes \citep[e.g.][]{Schrabback10,Blake11-baodistances,Riess11,Burenin12, planck2013-p11, Amati13}.
Investigating the uncertainties of $w$ measurements reveals that
distances to SNe are leading precision constraints.
The current constraints on a constant equation of state from a joint
fit of a flat wCDM cosmological model to the SN Hubble diagram and {\it Planck}
CMB measurements yields $w =-1.018 \pm 0.057 \mathrm{(stat+sys)}$ \citep{Betoule14}.

However, it is important to realise that in the quest for stricter
dark energy constraints, one should rely on several probes:
different probes face different parameter degeneracies and efficiently
complement each other; different probes are also subject to different systematic
uncertainties, and a cross-check is obviously in order for these
delicate measurements. Both arguments are developed in detail in
\cite{DETF06,ESO-ESA}.

When one constrains cosmological parameters from distance data,
increasing the redshift span of the data efficiently improves the
quality of cosmological constraints, and SN surveys are hence targeting the highest
possible redshifts. Cosmological constraints from SNe derive from comparing
event brightnesses at different redshifts. For precision cosmology, 
one should aim at
comparing similar restframe wavelength regions at all redshifts, so that
the comparison does not strongly rely on a SN model. When aiming at
higher and higher redshifts, ground-based SN surveys face two
serious limitations related to the atmosphere: at wavelengths redder
than $\sim$800~nm, the atmosphere glow rises rapidly in intensity;
this glow goes with large and time-variable atmospheric absorption
which makes precision photometry through the atmosphere 
above $1~{\rm \mu}$m very difficult.

Cosmological constraints from SN distances are currently dominated by
distances measured in the visible, mainly at $z\lesssim 1$,
(e.g. \citealt{Conley11,Scolnic14,Betoule14}). 
The
current sample of SN distances at $z>1$
is dominated by events measured with NIR instruments (NICMOS and WFC3) on the
HST \citep{Riess07,Suzuki12,Rodney12,Rubin13} and amounts to less than 40 such events. These NIR instruments have a small field of view compared to current
ground-based CCD-mosaics.
Extending the Hubble diagram of supernovae at
$z\gtrsim 1$ with statistics matching forthcoming ground-based samples
at $z\lesssim 1$ requires NIR wide-field imaging from space.

Euclid is an ESA M-class space mission, adopted in June 2012, which aims at
characterising dark energy, from two main probes \citep{EuclidRB}: the spatial
correlations of weak shear, and the 3-D correlation function of
galaxies. The latter allows one to measure in particular the evolution
of the expansion rate of the universe by tracking the BAO peak as a
function of redshift, while the study of the shear as a function of
redshift constrains both the expansion rate evolution and the growth
rate of structures. The growth rate of structures, and its evolution with
redshift can also be probed by extracting redshift space distortions
from the envisioned 3-D galaxy redshift survey. Measuring both the
expansion history and the growth rate evolution with redshift provides
a new test of general relativity on large scales because this theory
predicts a specific relation between these two aspects. Alternative theories
of gravity, which might be invoked instead of dark energy, predict
in general a different relation between growth of structures and expansion history
\citep[e.g.][and references
therein]{Lue04,Linder05,Bean07,Bernardeau11,Amendola13}.

Euclid will be equiped with a wide-field NIR imager and is hence well suited to
host a high-statistics high-redshift supernova programme, aimed at extending
the ground-based Hubble diagram beyond $z\sim 1$. This paper proposes 
such a SN survey and evaluates the cosmological
constraints it could deliver in association with measurements 
of distances to SNe at lower
redshifts with ground-based instruments. An earlier paper \citep[][A11 thereafter]{Astier11} 
aimed at designing a standalone space-based SN survey and suggested
a different route: it assumed that
a Euclid-like mission could be equipped with a filter wheel on its visible imager, which
is no longer a plausible possibility within the adopted mission constraints. 
However, some arguments developed in A11 still apply to the work presented here and we will
refer to this earlier study when applicable.

We present here a SN survey which addresses
systematic concerns and delivers valuable leverage on dark energy.
The plan of this paper is as follows: we will first discuss the
requirements of SN~Ia surveys for high-quality distances (\S\ref{sec:requirements}). We then describe
the salient points of our SN and instrument simulators
(\S\ref{sec:simulators}). The proposed surveys are described in
\S\ref{sec:sn-surveys}, and the assumptions regarding redshifts and typing
in \S\ref{sec:redshifts-typing}. The forecast methodology and the
associated Fisher matrix are the subjects of \S\ref{sec:methodology}. Our results are presented in \S\ref{sec:results}, and we explore alternatives to the baseline surveys in
\S\ref{sec:alteration}. In \S\ref{sec:des-wfirst}, we compare our findings to forecasts
for the SN survey projects within DES and WFIRST.
We discuss issues related to astrophysics of supernovae 
and their host galaxies in \S\ref{sec:astro-issues}.
The data set we propose to collect with Euclid allows a wealth of other
science studies, and we present a sample of those in \S\ref{sec:deep-vis-ir}.
We summarise in \S\ref{sec:summary}.

\section{Requirements for the supernova survey}
\label{sec:requirements}

Distances to SNe~Ia rely on the comparison of supernova fluxes at
different redshifts. The evolution of distances (up to a global
multiplicative constant) with redshift encodes the expansion history
of the universe. We will now discuss various aspects of the SN survey
design intended to limit the impact of systematic uncertainties.

One can summarise the current impact of systematic uncertainties on
SN cosmology \citep[][also known as
  SNLS3 ]{Guy10,Conley11,Sullivan11}: the photometric calibration
uncertainties dominate by far over other systematics, and contribute to the
equation of state uncertainty by about as much as statistics (see e.g. Table~7 from \citealt{Conley11}). For the latest SN+{\it Planck} results \citep{Betoule14},
the calibration uncertainties increase $\sigma(w)$ from 0.044 to 0.057. 
There are, however, ways to reduce the impact of calibration uncertainties
both in the survey design, and in the calibration scheme. Regarding
the latter, adding new calibration paths \citep{Betoule13} to the
classical path via the Landolt catalogue \citep[e.g.][]{Regnault09} already
reduced significantly the calibration uncertainty. Half of the current calibration
uncertainty is due to primary calibrators which is expected to decrease in the future.
The SNLS3 compilation
is dominated at high redshift by the SNLS sample, measured in the visible
from the ground using a camera that has limited sensitivity in its 
reddest band (i.e. the z-band). This specific feature affects the precision of 
SNLS distances at $z \gtrsim 0.8$, as discussed below. We discuss now how
the SN survey design can mitigate calibration uncertainties.

\subsection{Wavelength coverage}
\begin{center}
\begin{figure}[ht]
\includegraphics[width=\linewidth]{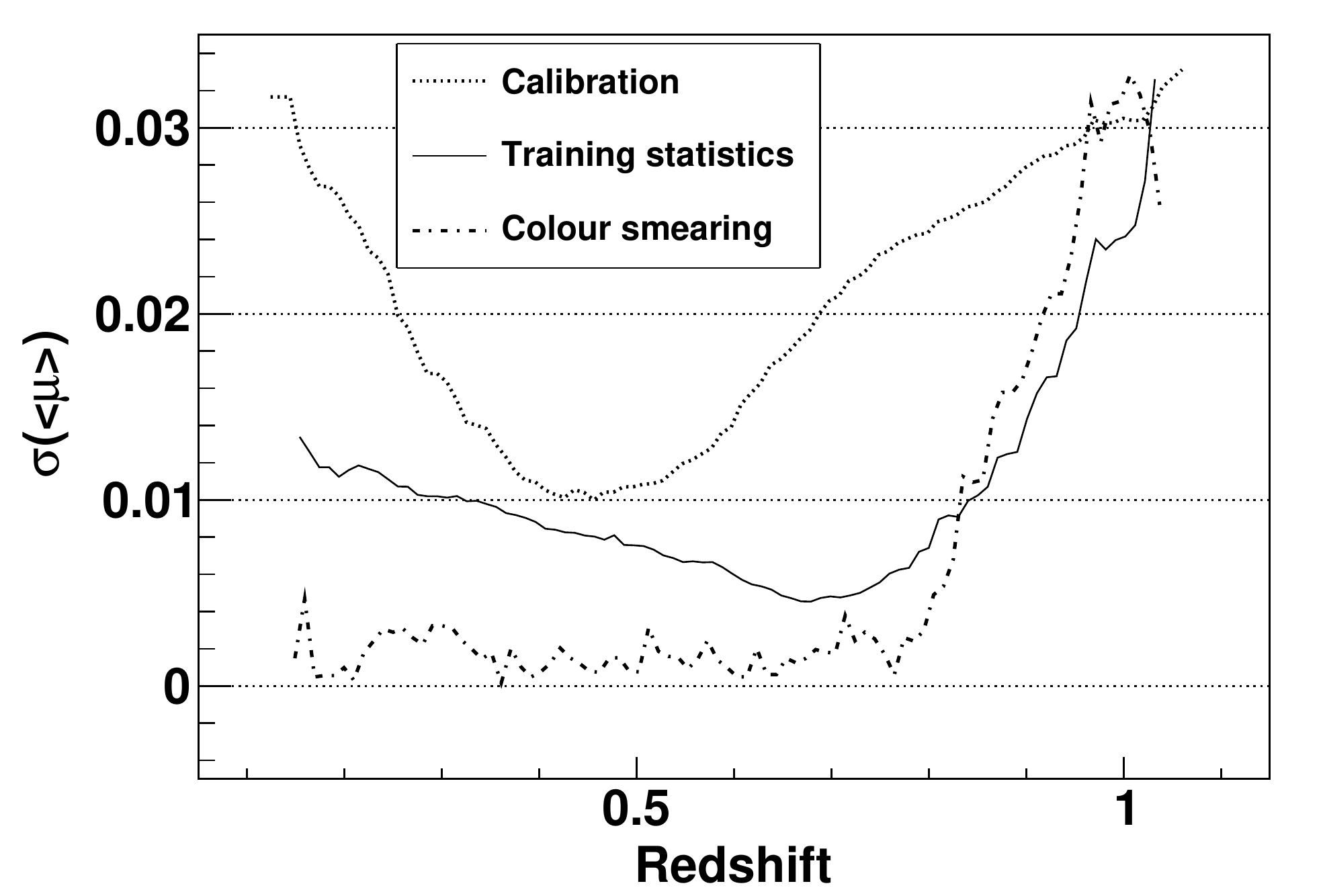}
\caption{Contribution of various sources to correlated uncertainties,
  averaged over sliding $\Delta z$=0.2 bins for the SNLS3 analysis
  (data from \citealt{Guy10}).
  ``Colour smearing'' refers to the effect of uncertainties 
  of the band-dependent residual scatter model (see \S\ref{sec:sn-simulator}). The steep
  increase at high redshift of this contribution  and of that from SN model training
statistics
   are both due to those events being measured in
  bands bluer in the rest-frame than the lower redshift events. 
We note that these two contributions are indeed going down with sample size.
\label{fig:combined_uncertainty}}
\end{figure}
\end{center}

If fluxes of SNe at different redshifts are measured at different {\it
  restframe} wavelengths, one has to rely on some modelling of the
spectrum of SNe in order to convert relative
fluxes to relative distances. Distances relying on such a model are
affected by systematic and statistical uncertainties from this model,
 correlating all
events at the same redshift. This effect is illustrated in the case of the
SNLS survey by the Fig. \ref{fig:combined_uncertainty}, where one can
see that at the high-redshift end, uncertainties unrelated to the
measurement itself become important, especially because they are
common to all events. Because of the low sensitivity of the imager 
in $z$ band, these high redshift events are effectively measured in bluer
restframe bands than events at lower redshifts, which makes their distances sensitive to
statistical and systematic uncertainties of the SN model. 
This SN model always derives from a training sample and inherits
all uncertainties affecting this training sample. 
In particular, the calibration uncertainties affecting the SN model training
sample propagate to these distances to high-redshift events
measured in restframe bands extending bluer than $U$. So, a strategy 
requiring that all events be measured in similar restframe bands reduces
the impact of SN model uncertainties on distances. We
propose below a quantitative implementation of this requirement.

\subsection{Amplitude, colour, and distance uncertainties}
The signal-to-noise ratio of the photometric measurements affects the precision of
distances, but at some point, distances will not significantly benefit from deeper
exposures. We discuss here current intrinsic limitations 
of supernova distances as well as how measurement precision
contributes to distance precision.

SNe~Ia exhibit some variability both in light curve shape and
colour, both correlated with brightness \citep[e.g.][and references
  therein]{Tripp99} and most SN distance estimators rely
in some way on these brighter-slower and brighter-bluer relations.
A common way of parametrising a distance modulus $\mu\equiv 5 log_{10}(d_L)$, accounting
for these relations is 
\begin{equation}
\mu = m_B^* + \alpha(s-1) -\beta c - {\mathcal M},
\label{eq:distance-estimator}
\end{equation}
where $m_B^*,s,c$ are fitted SN-dependent parameters. $m_B^*$ denotes
the peak brightness in restframe $B$ filter, $s$ is a stretch factor
describing the light curve width (or decline rate), and $c$ is a
restframe colour most often chosen as $B-V$ evaluated at peak
brightness.  $\alpha$, $\beta$ and ${\mathcal M}$ are global
parameters derived from data (and subsequently marginalised over),
typically by minimizing the distance scatter. They do not convey
cosmological information, but rather parametrise the brighter-slower,
brighter-bluer and intrinsic brightness of SNe. For each event, the $m_B^*,s,c$ parameters are derived
from a fit of a SN model to the measured light curve points, in at
least two bands, if colour is to be measured. The $m_B^*$ and $c$ parameters, which mainly
  determine the distance precision, are derived from amplitudes of
  light curves in different bands, where ``amplitude'' refers to some
  brightness indicator (e.g. the peak brightness) derived from the
  light curve in a single band. We will now discuss the requirements
  on the quality of photometric measurements, and express those using
  amplitude precision.

\begin{center}
\begin{figure}[h]
\includegraphics[width=\linewidth]{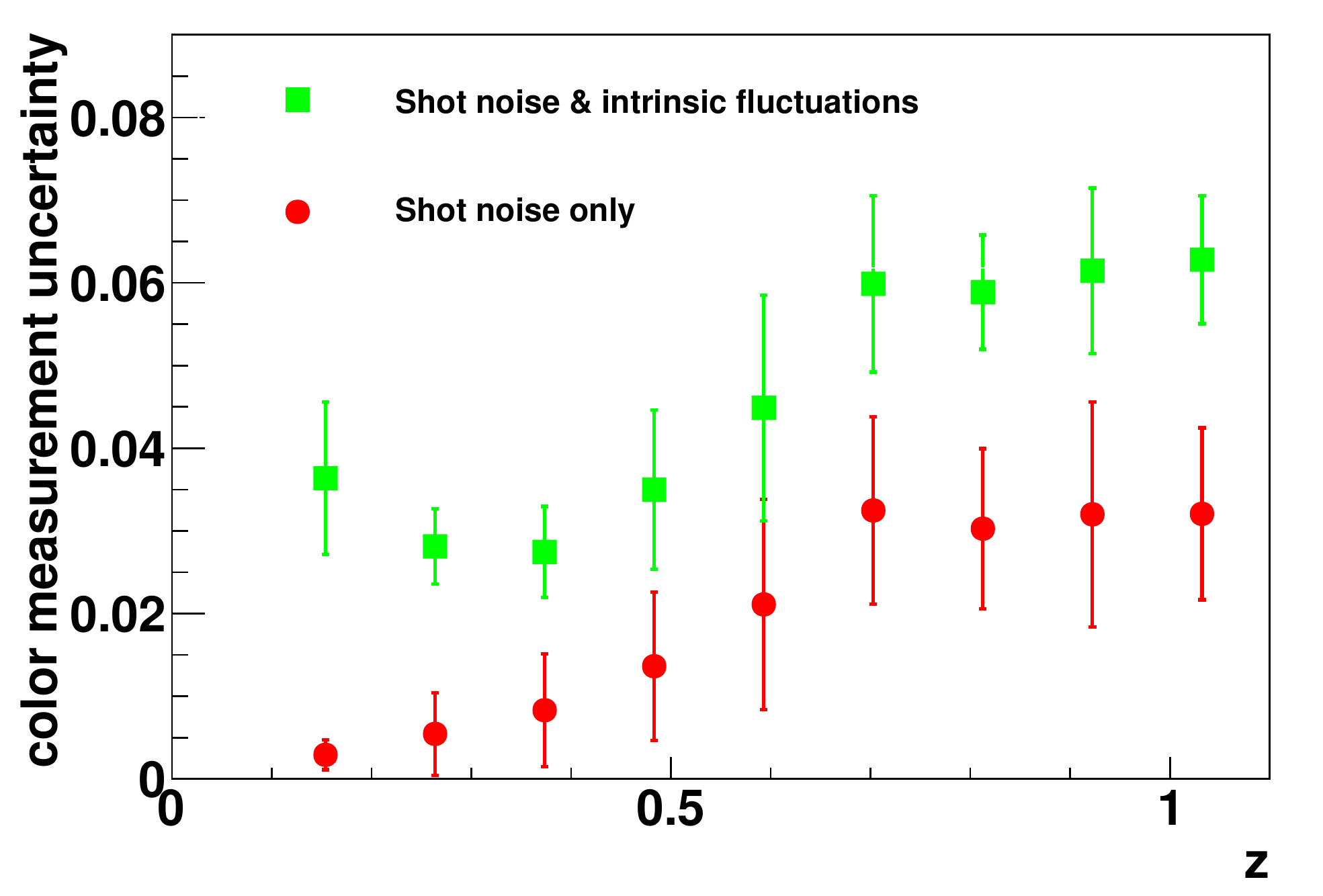}
\caption{Measurement uncertainty of the $c$ parameter in the SNLS
  survey as a function of redshift, for events spectroscopically
  identified. Solid circles show the contribution of the source shot noise alone,
and the squares include intrinsic fluctuations from event to event (also called colour smearing). 
At $z>0.7$, the shot noise contribution 
  becomes essentially constant because the colour measurement
  relies on bluer and bluer restframe bands, which are more and more
  sensitive to colour changes. This might look favourable,
  but accounting for intrinsic fluctuations from event to event (squares),
  very large in the UV, swamps this benefit. (Data
  obtained from fitting light curves from \citealt{Guy10}).
\label{fig:snls-color-error}}
\end{figure}
\end{center}

The contribution of the $s$ uncertainty to the $\mu$ uncertainty is 
sub-dominant for light curves spanning at least $\sim$30
restframe days. On the contrary, since $\beta$ turns out to be larger
than 1 (for $c=B-V$ restframe, $\beta_{B,V}$ is indeed measured to be
above 3, see e.g. \citealt{Guy10}), the $c$ measurement uncertainty
drives the distance measurement uncertainty.  Since the observed
scatter of SNe distance moduli (given by Eq.~\ref{eq:distance-estimator})
around the Hubble diagram is at best about 0.15 mag, $c$ measurement
uncertainties above $\sim 0.04$ mag will start to contribute significantly to
the distance uncertainty. Fig. \ref{fig:snls-color-error} shows that
the SNLS survey is within this bound up to $z=1$. This performance is
however obtained on a sample that is effectively flux-selected by
spectroscopic identification, and that relies on the $r$ band to measure
colour at the highest redshifts. This restframe UV region is affected
by large fluctuations from event to event (Fig. 4 of \citealt{Maguire12}, Fig. 8 of
\citealt{Guy10}). Worse, Fig. 4 of \cite{Maguire12} may suggest
an evolution with redshift of the flux at wavelengths shorter than 320 nm. 
So, we give up the rest-frame UV region
by requiring that filters with central wavelength below 380 nm
in the rest-frame are not used for distances. Amplitudes of light curves in
the BVR rest-frame region measured to a precision of 0.04 mag deliver 
a colour precision 
of about 0.045 mag with two bands, and better than 0.03 mag with 3 bands. We note that
measurements in the rest-frame UV, even if not used for distances,
are still available for photometric identification. Measurements
at $\sim$280 nm (rest frame) are available in certain redshift ranges, and can be used as a
possible control of evolution of supernovae, as discussed in \S\ref{sec:spectro-UV}.

When a sizable fraction of the SN Ia population is lost at the
high-redshift end of the Hubble diagram because of flux selection, one
has to simulate the unobserved events to correct for the bias of the
observed sample. This procedure aims at compensating for the so-called
Malmquist bias, but the uncertainties of such a procedure \citep[see
e.g.][]{WoodVasey07,Kessler09,Perrett10,Conley11,Kessler13} limit the
usefulness of an incomplete high redshift sample. On top of possible
systematics, there is a statistical price to pay: an incomplete high
redshift sample is on average bluer than the whole population, and
induces correlations between $\beta$ (Eq.~\ref{eq:distance-estimator})
and cosmological parameters\footnote{From the distance modulus
  definition (Eq.~\ref{eq:distance-estimator}), one can infer that if
  the average colour $\langle c \rangle$ is independent of z, the
  average distance modulus $\langle \mu \rangle$ does not depend on
  $\beta$ and hence estimates of cosmological parameters and $\beta$
  are independent. In \cite{Kessler13} (\S 6.4), it is shown that the
  value of $\beta$ influences both the evaluation of Malmquist bias
  and distance moduli in ways which tend to cancel each other on
  average. The statistical coupling between cosmological parameters
  and $\beta$ however remains.}  which degrade the quality of
cosmological constraints. Conversely, if the SN colour distribution of
the cosmological sample is the same at all redshifts, a wrong $\beta$
or even an inadequate form of the colour correction affects SNe at all
redshifts in the same way, and hence does not alter the average
distance-redshift relation. So, all efforts need be made to retain a
very large fraction of the population at the highest redshift. Since
high-redshift red SNe are very faint and thus missing from SN samples,
one can eliminate the potential bias by ignoring red events at all
redshifts. The analyses typically reject both blue and red events
beyond 2.5 to 3$\sigma$ (see e.g. \citealt{Kessler09,Conley11}) from the
mean of the restframe $B-V$ distribution and the statistical cost is at the 
few percent level.

\subsection{Light curve measurement precision requirements}
\label{sec:lc-requirements}
\begin{center}
\begin{figure}[ht]
\includegraphics[width=\linewidth]{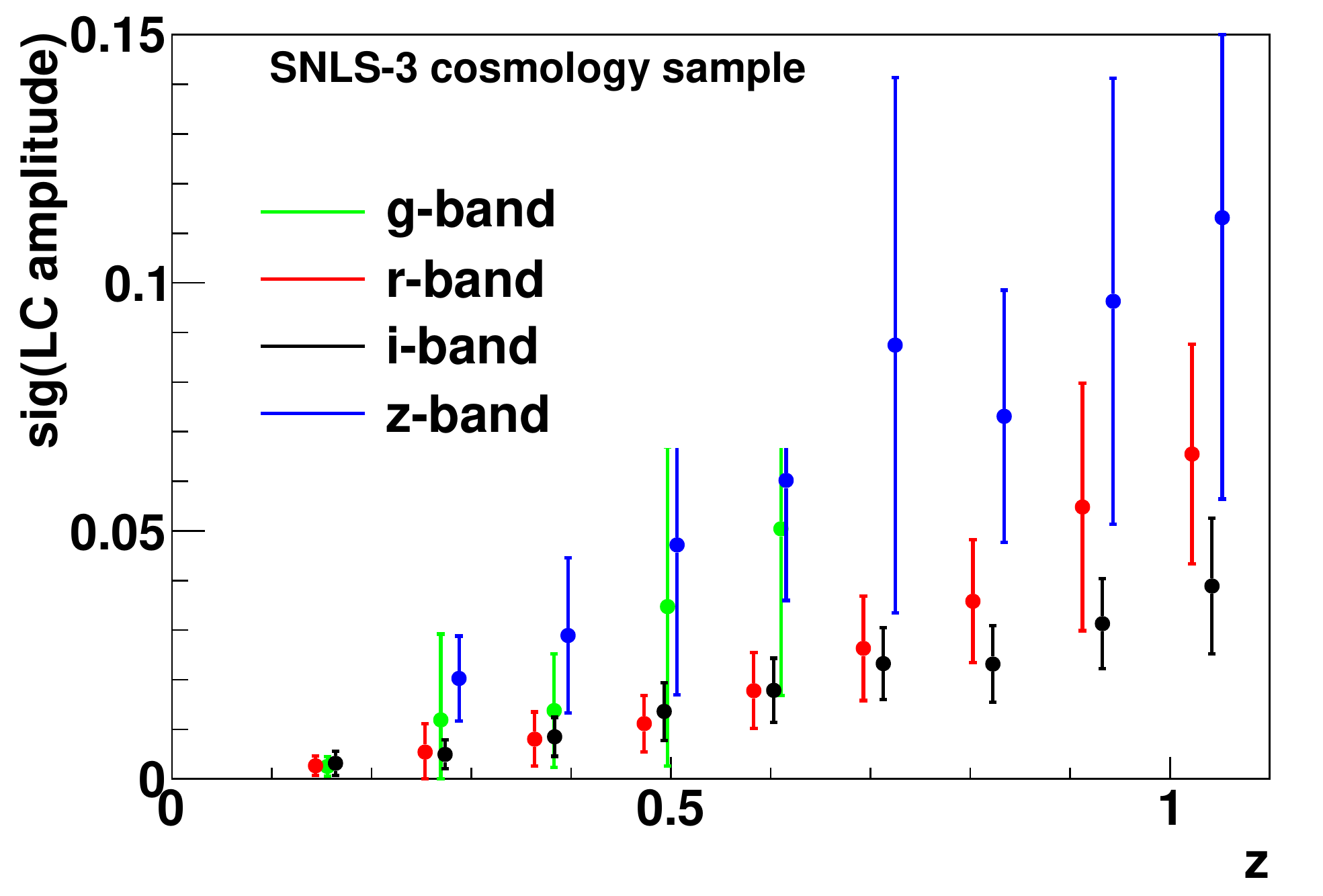}
\caption{Measurement uncertainties of fitted amplitudes of SNLS light curves,
propagating shot noise.
The $i$-band precision is below 0.03 mag up to $z=1$, as well as the r-band up to 
$z \simeq 0.75$. SNLS observations rely on thinned CCDs with a low QE in z-band.
This band is thus shallow and hence has a small weight in distances to high-redshift events.
(Data from fitting light curves from \citealt{Guy10}).
\label{fig:snls-amp}}
\end{figure}
\end{center}

We propose the following quality requirements for photometric measurements
of SNe~Ia aimed at deriving distances:
\begin{enumerate}
\item We express the quality of light curve measurements from the r.m.s
uncertainty of their fitted amplitude. Our goal is to secure two bands measured to
a precision of 0.04 mag and a third band to 0.06 mag.
 {\it Rationale :} this ensures a colour measured to 0.03 mag, such that
  the colour uncertainty is 
  sub-dominant in the distance uncertainty. 
As long as measurements meet this
quality, there are no
detection losses, because detection and photometric measurements are carried out
from the same images. By discarding events at redshifts that do not meet these
quality requirements, we effectively construct redshift-limited surveys.
\item Do not use filters with central wavelength below 380~nm in the
  restframe. {\it Rationale:} SNe~Ia have large dispersions in the UV,
and there are indications of evolution below 330~nm.
\item Derive distances from most similar restframe regions
  at all redshifts. To this aim, we only consider filters with central wavelengths $380<\lambda<700$ nm.
{\it Rationale:} reduce dependence on SN model
  and its associated systematic
  (e.g. calibration of the training sample) and statistical
  uncertainties.
\item Measure light curves over [-10,+30] restframe days from maximum
  light.  {\it Rationale :} measure light curve width in order to account for the 
  brighter-slower relation, and provide light curve shape information 
  for SN typing. Compare rise and decline rates across redshifts
  for evolution tests.
\end{enumerate}
These requirements will be used as guidelines for the SN survey designs in \S \ref{sec:sn-surveys}.
Fig.~\ref{fig:snls-amp} shows that the SNLS observations meet these requirements
up to $z=0.65$; they fail at higher redshifts because of the modest sensitivity in z-band (the CCDs of Megacam \citep{MegacamPaper} are optimised for blue wavelengths).
An imager equipped with deep-depleted thick CCDs can meet our requirements
up to $z\simeq 0.95$, acquiring deep enough y-band data, and with exposures 
significantly deeper than SNLS.
The strategy proposed for DES in \cite{Bernstein12} does not provide
three bands redder than 380~nm at $z\gtrsim 0.68$, because it does not plan on using the low-efficiency
y band.

Euclid hosts a visible imager, called VIS, equipped with a single broad
  band $500\lesssim\lambda\lesssim 950$~nm, in order to maximise the
  S/N of galaxy shape measurements. Such a band corresponds to merging two to
  three regular broadband filters.  The requirements above exclude
  using this band for measuring distances to SNe at $z \gtrsim 0.5$,
  because at higher redshifts, it includes too blue rest-frame
  regions. More generally, our requirement that measurements are
  similar across redshifts excludes an observer band much wider than
  the others. However, deep Euclid visible data of the SN hosts will
  be valuable for other reasons, discussed in \S~\ref{sec:deep-vis-ir}.

\subsection{Cadence of the survey}
In the above requirements, we have not discussed the sampling cadence
along the light curves because we have expressed the depth requirement
directly on the fitted light curve amplitude (point 1). If an observing
cadence meets this requirement, visits twice as frequent integrating half the
exposure time will not change significantly
the precision of the fitted amplitude. As a baseline, we adopt in what follows a
four-day cadence in the observer frame, because this is more than
adequate to sample light curves of high-redshift supernovae and allows one to efficiently
study faster transients. We could measure distances to SNe~Ia using 
a somehow slower cadence, but with accordingly deeper exposures at each visit.

\section{Instrument and supernova simulators}
\label{sec:simulators}

\subsection{Instrument simulator}
\label{sec:instrument}

In its current design, Euclid is equipped with a visible and a NIR
 imager \citep{EuclidRB}.  The latter also has a slitless
spectroscopic mode but what we will discuss here does not rely on this
capability, mainly because high-redshift SNe are too faint for
slitless spectrocopy on Euclid to deliver a usable signal. We do not
rely either on the visible imager for measuring distances, as mentioned
above.
Therefore, the SN observations we are going to
discuss rely solely on the NIR Euclid imager.

\begin{figure}
\centering
\includegraphics[width=\linewidth]{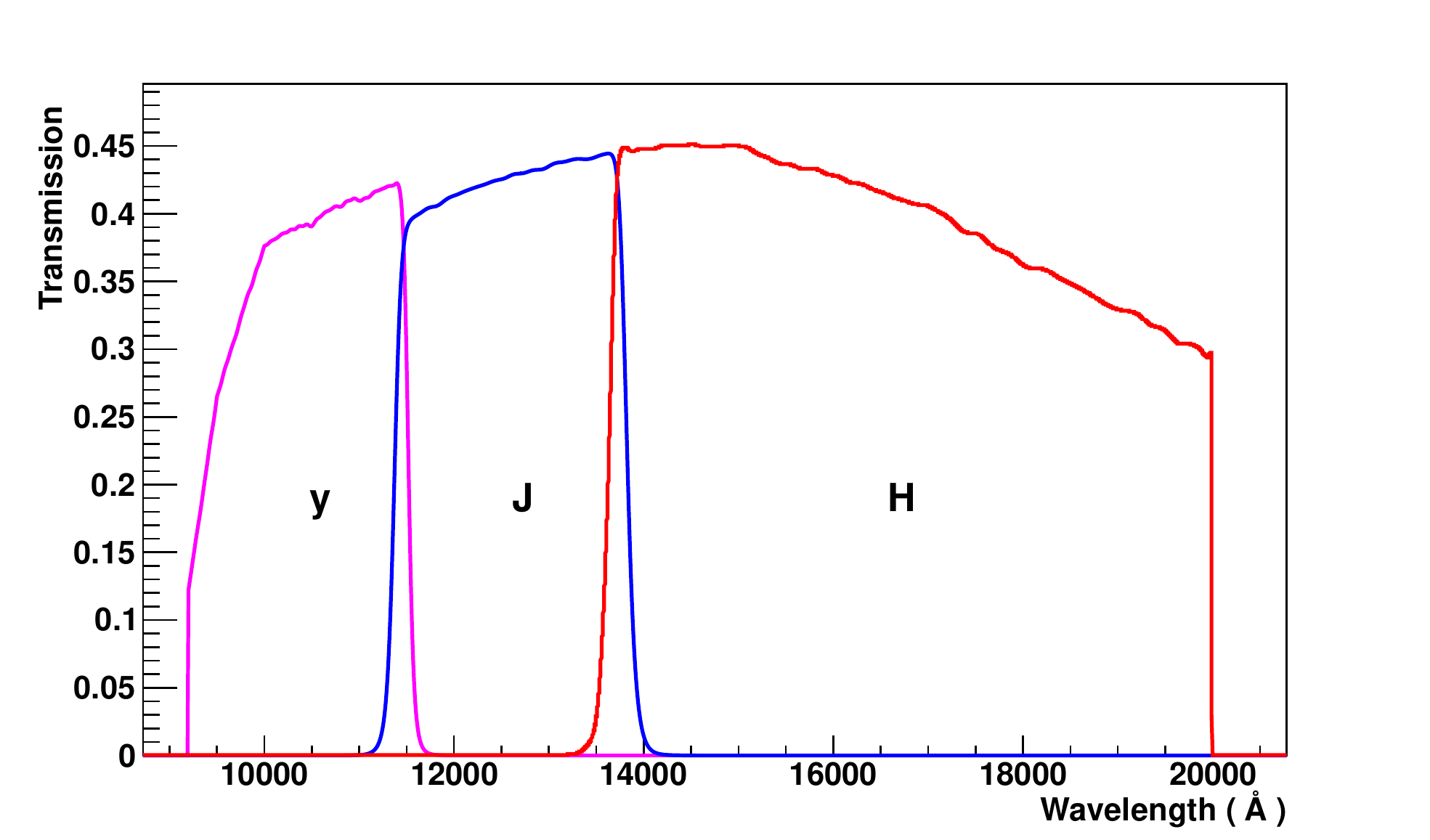}
\caption{Overall transmission of the 3 bands of the Euclid 
NIR imaging system, in its current design. The H filter
red cutoff has been pushed to 2 $\mu$m  compared to earlier designs.
The cut-on of the y filter is determined by the dichroic that splits
the beam between visible and NIR instruments. 
\label{fig:euclid-bands}
}
\end{figure}

In  order to assess the cosmological performance of possible
surveys, we simulate SN observations in Euclid and 
other imaging instruments. The first step is to evaluate
the precision of photometric measurements. 
For a given SED, observing setup and observing strategy, our simulator computes the
expected flux and evaluates the flux uncertainty assuming measurements
are carried out using PSF photometry for a given sky
background and detector-induced noise, and accounts
for shot noise from the source; this calculation is described in appendix~\ref{sec:point-source-photometry}. For Euclid's NIR
imager, we use PSFs derived from full optical simulations (including diffraction)
and detector characteristics\footnote{We are in debt to R. Holmes for providing
us with the PSFs, transmission curves displayed in Fig.~\ref{fig:euclid-bands} and sensor characteristics which allowed
us to simulate NIR imaging with Euclid.}. These optical simulations were used to define the
exposure times for NIR imaging in the Euclid observing plan for its core science.
The most important parameters of our Euclid NIR imager simulator are :
\begin{itemize}
\item a mirror area of 9300 cm$^2$,
\item a read-noise of 7 electrons,
\item a dark current of 0.1 electrons/pixel/s,
\item pixels subtend 0.3\arcsec\  on a side,
\item and the imager covers 0.5 deg$^2$ on the sky.
\end{itemize}
This NIR imager has 3 bands (named y, J and H) roughly covering the $[$1-2$]\ {\rm
  \mu}$m interval.  The overall transmission of the imager bands
(accounting for all optical transmissions and quantum efficiency of
the sensors) are shown in Fig.~\ref{fig:euclid-bands}. The important
parameters of the simulated photometry bands are provided in Table~\ref{tab:sky-zp}.

\begin{table}[h]
\caption{Characteristics of the Euclid bands simulated for the high-redshift survey.}
\begin{center}
\begin{tabular}{|c|ccccc|}
\hline
    & $\bar{\lambda}$ & $S_{45}$ & $S_{15}$ & ZP & NEA\\
band & (nm) & \multicolumn{2}{c}{(AB/$\mathrm{arcsec^2}$)} & ($\mathrm{m_{AB}\ for\ 1e^-/s}$) & (arcsec$^2$) \\
\hline
y & 1048 & 22.47 & 22.75 & 24.03 & 0.56 \\
J & 1263 & 22.44 & 22.72 & 24.08 & 0.61 \\
H & 1658 & 22.31 & 22.60 & 24.74 & 0.77 \\
\hline
\end{tabular}
\end{center}
\tablefoot{Columns include: central wavelength, sky brightness (in AB
  magnitudes/${\rm arcsec^2}$) at two separations from the ecliptic poles (45$^\circ$ and 15$^\circ$),
  zero-points (for AB magnitudes and fluxes in~${\rm e^-/s}$), and
  Noise Equivalent Area (NEA) of the PSF (defined by Eq.~\ref{eq:NEA-definition}). This is the area over which
  one integrates the sky background fluctuations when performing PSF
  photometry for faint sources, accounting for pixelisation at 0.3''/pixel. 
  The reported NEA 
  values were averaged over source position within the central
  pixel.
\label{tab:sky-zp}}
\end{table}

We have used the zodiacal light models in space from \cite{Leinert98},
more precisely the angular dependence from their Table~16, and the spectral
dependence from their Table~19. The zodiacal light intensity depends on the
ecliptic latitude because of the albedo of solar system dust, and the
darkest spots are the ecliptic poles. Our Table~\ref{tab:sky-zp}
presents sky brightnesses at two ecliptic latitudes. 
The brightest one, $S_{45}$ refers to $45^\circ$
from the ecliptic pole where we assumed a zodiacal light flux density
normalised to $7.54\ 10^{-19} \ {\rm ergs/(cm^{2}s\AA \ arcsec^{2})}$
at 1.2 ${\rm \mu}$m. With this value, our simulator derives 5~$\sigma$
limiting AB magnitudes of 24.02, 24.03, and 23.98 for three exposures of
79, 81 and 48 s in y, J and H respectively, assuming PSF photometry is
carried out. These values compare very well to the limiting magnitudes
of 24.00 (set by scientific requirements, see \citealt{EuclidRB}) found by the instrument development team, who indeed derived
the above exposure times of the ``Euclid standard visit'' that deliver this
sensitivity.

Fields selected to monitor SN light curves have to be observable over
long periods of time, and the Euclid spacecraft design imposes that they are located near
the ecliptic poles. We will hence use in what follows the $S_{15}$ sky
intensities from our Table~\ref{tab:sky-zp} which apply at 15$^\circ$
and closer to the ecliptic poles. Our 5~$\sigma$ limiting magnitudes
for 3 standard Euclid exposures (79,81,48 s in y,J,H) then become 24.05, 24.07 and 24.03 in
y, J and H respectively, i.e. they are improved by $\sim$0.04 with respect to
45$^\circ$ from the ecliptic poles. The improvement with decreasing
sky background is modest because read noise contributes 
$\sim$ 60 \% to the total noise of the NIR standard Euclid exposures.

\subsection{Impact of finite reference image depth}
\label{sec:ref-depth}
Supernovae photometry is obtained by subtracting images without the
supernova (deemed the reference images) from images with the
supernova. Since the same SN-free images are subtracted from all light
curves measurements, the SN fluxes along the light curve are
positively correlated, and have a larger variance than the fluxes 
before subtraction. This correlation and extra variance both vanish 
for an infinitely
deep reference image, but since we will not have an infinitely deep
reference, the precision of light curve amplitude measurements is
degraded with respect to this ideal case. We detail in appendix~\ref{app:ref-depth} the computation of
the effect, and will come back later to its practical implications.

Beyond the contribution to shot noise, differential photometry
  might also contribute to systematic uncertainties, especially in the
  context of ground-based image sets with sizable variations of image
  quality. Tests on real images from a ground-based SN survey 
\citep{AstierPhotom13} show that it is possible to obtain systematic residuals 
below 2 mmag,
hence negligible compared with calibration uncertainties. The same
tests show that the observed scatter of SN measurements 
follows the expected contributions from shot noise.

\subsection{Supernova simulator \label{sec:sn-simulator}}

To simulate SNe~Ia, we primarily made use of the SALT2 model
\citep{Guy07,Guy10}. This model is a parametrised spectral sequence,
empirically determined from photometric and spectroscopic data.
We also made use of the brighter-slower and brighter-bluer 
relations determined from the SNLS3 SN sample \citep{Guy10},
and the average absolute magnitude $M_B = -19.09 + 5 log_{10} (H_0/70 km/s/Mpc)$
in the Landolt (i.e. Vega) system. Because of limitations of its training sample,
SALT2 does not cover restframe wavelengths redder than 
800 nm. 

SALT2 parametrises events with 4 parameters: a date of maximum light
(in $B$-band) $t_0$, a colour $c$, a decline rate parameter $X_1$ and an overall
amplitude $X_0$. The latter is often expressed as $m_B^*$, the
peak magnitude of the light curve in
the redshifted $B$-band. Given these parameters, a redshift and a
luminosity distance, we can evaluate fluxes of the SN in the observer
filter at the required phase, and evaluate the uncertainty of the
measurement, for the adopted instrumental setup and given observing conditions. Varying the cosmology only
alters $X_0$ (or $m_B^*$), and in the simulations, we have assumed
that the current uncertainties on the expansion history are now small
enough to ignore the changes of measurement uncertainties when varying
the cosmology.

   SALT2 does not assume any relation between brightness and redshift.
In the training process, the $X_0$ of events are nuisance parameters.
This allows one to decouple distance estimation from light curve
fitter training, and more importantly to train the light curve fitter
using data at unknown distance. Thus, 
the SALT2 trainings \citep{Guy07,Guy10} use a mixture of 
nearby events (including
very nearby events where the redshift
is a poor indicator of distance) and well-measured SNLS events.
Since the statistical uncertainty of the model eventually
contributes to the cosmology uncertainty, one has to minimise
the former. In what follows, we will emulate the LC fitter training
in order to incorporate the uncertainties that arise from this process
into the cosmology uncertainties. We note that the light curve 
fitter training suffers from both statistical uncertainties 
(from the size and quality of the training sample) and from systematic
uncertainties (typically the photometric calibration).

SALT2 is not a perfect description of SNe~Ia, and there remain some
variability of light curves around the best fit to data, beyond
measurement uncertainties. This scatter depends on the adopted supernova model
and was determined for SALT2 in \cite{Guy10}. The residual scatter is
described there as a coherent move around the average model of all light curve points 
of each band of each event, and it is found to depend on the restframe
central wavelength of the band. This scatter is measured to about
0.025 mag rms in BVR-bands and increases slowly towards red and very
rapidly in the UV (Fig. 8 from \citealt{Guy10}). This
scatter (coined ``colour smearing'' in \citealt{Kessler09}) is accounted for
in the simulation, and causes the difference between the two sets of
points in our Fig.~\ref{fig:snls-color-error}. 
\cite{Kessler13} considers other colour smearing
models than the SALT2 one and finds that this does not have a
dramatic effect on the recovered cosmology. We note that
the sample size we are considering in this paper will allow us to
considerably narrow down the range of acceptable colour smearing models.

For the rate of SNe, we use the volumetric rate from \cite{Ripoche08}
\begin{equation}
R(z) = 1.53\ \ 10^{-4} \left [ (1+z)/1.5 \right ]^{2.14} h_{70}^3\  \mathrm{Mpc^{-3}\  yr^{-1}} \label{eq:rate},
\end{equation}
where years should be understood in the rest frame.  Since these
measurements stop around $z=1$, rates at higher redshifts were
assumed to become independent of $z$. These rates compare well with the
determination from \cite{Perrett12}.  The rates proposed in
\cite{Mannucci07} (accounting for events ``lost to extinction'') yield
a SN count (to $z=1.5$) $\sim$ 25 \% larger than our nominal assumption,
with a similar redshift distribution. 
There are determinations of SN~Ia rates
at $z>1$ from the Subaru deep field (\citealt{Graur11}), and from the
CLASH/Candels survey (\citealt{Graur14,Rodney14}), which are compatible
with each other (see e.g. fig 1. of \citealt{Rodney14}), and show
that our assumption of rates flattening at $z=1$ is likely conservative
at the 20 to 30 \% level. We will discuss later (\S \ref{sec:redshifts})
other sources of uncertainty affecting the expected number of high-redshift 
events and will eventually derive how the cosmological precision depends on
event statistics (\S \ref{sec:alteration}).
The redshift distribution
of simulated events accounts for edge effects, i.e. we reject events
at the beginning or the end of an observing season which do not have
the full required restframe phase coverage.

The supernova simulation generates light curves in the
user-required bands, at the user-required cadence, on a regular
(redshift, colour, stretch) grid. For each band of each event, we
evaluate the peak fluxes and the weight matrix of the four event
parameters, by propagating the measurement uncertainty of all
measurement points in this band, accounting for the effect of finite
reference depth (Eq.~\ref{eq:finite_ref_depth}). 
These peak flux values and weight matrices are used 
by a global fit (\S \ref{sec:methodology} below) which
will weight these events according to their redshift, colour and
decline rate using measured distributions from \cite{Guy10}. The event weight also depends on the redshift-dependent
SNe~Ia rate (Eq.~\ref{eq:rate}), the edge-effect corrected survey duration,
and the survey area. 
The colour smearing is accounted for during the global fit.

\section{Supernova surveys\label{sec:sn-surveys}}
The SNLS survey has delivered its three-year sample,
together with a cosmological analysis gathering the high quality SN
sample and accounting for sytematic uncertainties
\citep{Guy10,Conley11,Sullivan11}. This compilation amounts to about 500
well-measured events, and will grow to about twice as much when SNLS and SDSS
release their full samples, and gathering the nearby samples ($z<0.1$)
that appeared recently \citep[e.g][]{Stritzinger11,Hicken12,Silverman12}. 
Pan-STARSS1 has recently delivered a first batch of 112 distances to SNe Ia
at $0.1 \lesssim z \lesssim 0.6$ \citep{Scolnic14, Rest14}, corresponding 
to 1.5 y of observations. The 
next significant increase in statistics is expected from the Dark Energy 
Survey (DES), which aims at delivering $\sim 3000$ new events in a
5-year survey extending to $z\sim 1.2$ \citep{Bernstein12}, to which 
we compare our proposal in \S\ref{sec:des-wfirst}. To make significant
improvements, a SN proposal for the next decade should target
at least $10^4$ well-measured events and should aim at significantly 
increasing the redshift lever arm.  

\subsection{High-z SN survey with Euclid: the DESIRE survey}

As discussed in the introduction, measuring accurate distances
  to SNe at $z>1$ requires to observe from space in the NIR. With its
  wide-field NIR capabilities, Euclid offers a unique opportunity to
  deliver a large sample in this redshift regime. In this section, we
  present the DESIRE survey (Dark Energy Supernova Infra-Red
  Experiment) which will be a dramatic improvement in the number of
  high quality SNe~Ia light curves at redshifts up to 1.5.

Euclid observing time will be mostly devoted to a wide survey of 15 000 deg$^2$,
with a single visit per pointing \citep{EuclidRB}. Each single visit consists
of 4 exposures for simultaneous visible imaging and NIR spectroscopy,
and 4 NIR imaging exposures of 79, 81 and 48~s in y, J and H respectively.
We refer to this set of observations as the ``Euclid standard visit''.
The Euclid observing plan also makes provision for deep fields, which consist of repeated
standard visits, in particular in order to assess the repeatability of
measurements from actual repetition rather than from first principles. We 
attempted to assemble a SN survey from these repeated standard visits and failed
to find a compelling standalone SN survey strategy. Our unsuccessful attempts are described
in appendix \ref{sec:SN-in-deep}.

Since we aim at measuring 3 bands per event, and require that these 3 bands
map similar restframe spectral regions at all redshifts, we need to observe in more
than 3 observer bands in order to cover a finite redshift
interval. The obvious complement to Euclid consists of $i-$ and
$z-$bands observed from the ground. We identify at least three
facilities capable of delivering these observations: LSST \citep[8~m,][]{LivingLSST}, the Dark
Energy Camera (DECam) on the CTIO Blanco \citep[4~m,][]{DECAM-SPIE}, 
and Hyper Suprime Cam (HSC) on the Subaru \citep[8~m,][]{HSC-SPIE}. The most
efficient of these three possibilities is LSST; HSC would require
about 3 times more observing time than LSST while DECam would
require about 10 times more. While these are all plausible options, 
we consider LSST to be the most natural partner and we chose it 
to illustrate the DESIRE survey in the remainder of this paper.

\begin{figure}[h]
\includegraphics[width=\linewidth]{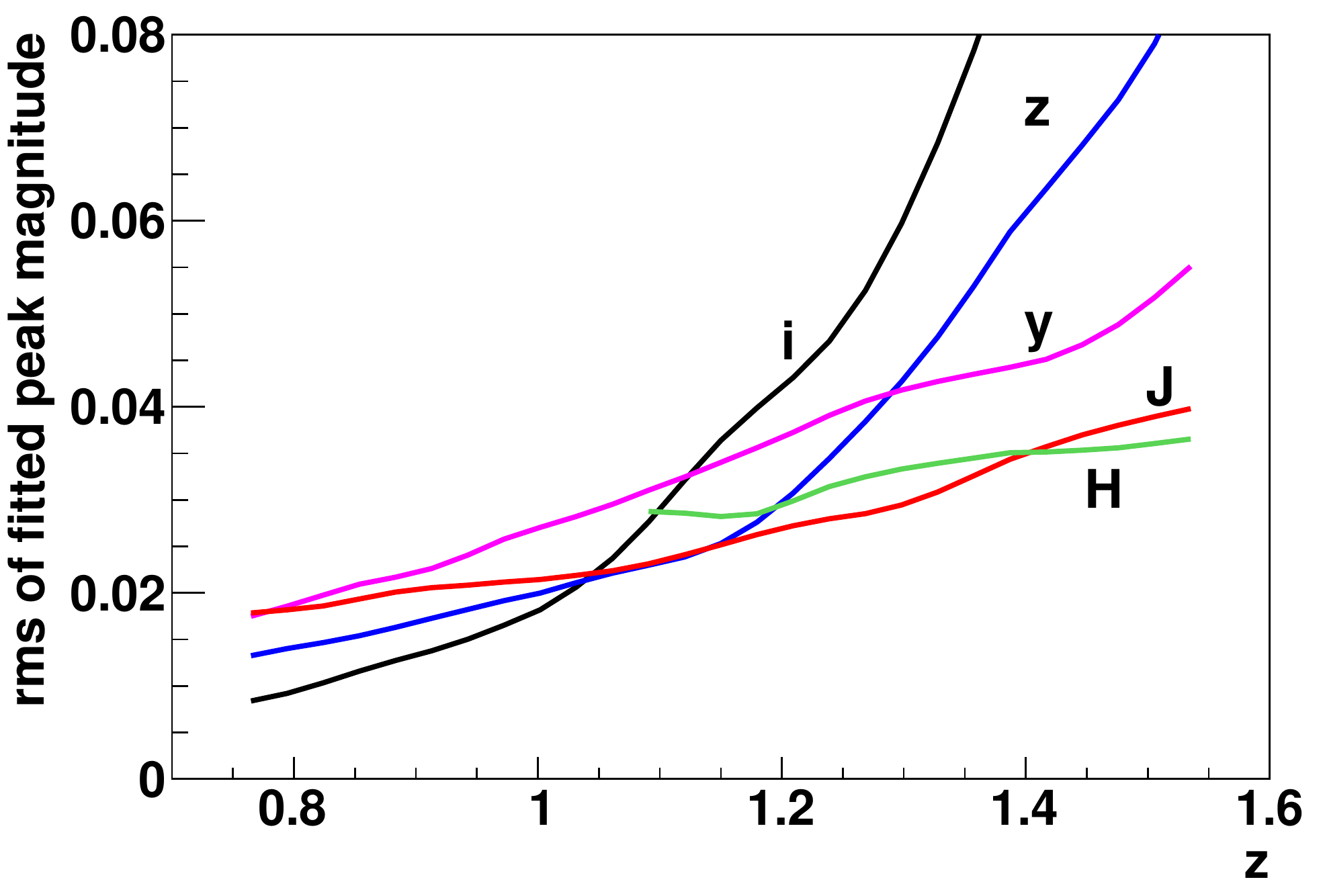}
\caption{Precision of light curve amplitudes as a function of redshift
  for the 5 bands of the DESIRE survey, assuming a 4-day cadence with
  the exposure times of Table~\ref{tab:joint-exp-times}. To fulfill
the requirements in \S\ref{sec:lc-requirements}, $i$-band is
  used up to $z=1$, $z$-band up to $z=1.2$, and distances at $z=1.5$ rely
  mostly on J- and H-band. For y, J and H bands, these calculations
  assume a reference image gathering 60 epochs in Euclid.
\label{fig:jointsurvey_reso}}
\end{figure}

\begin{figure}
\includegraphics[width=\linewidth]{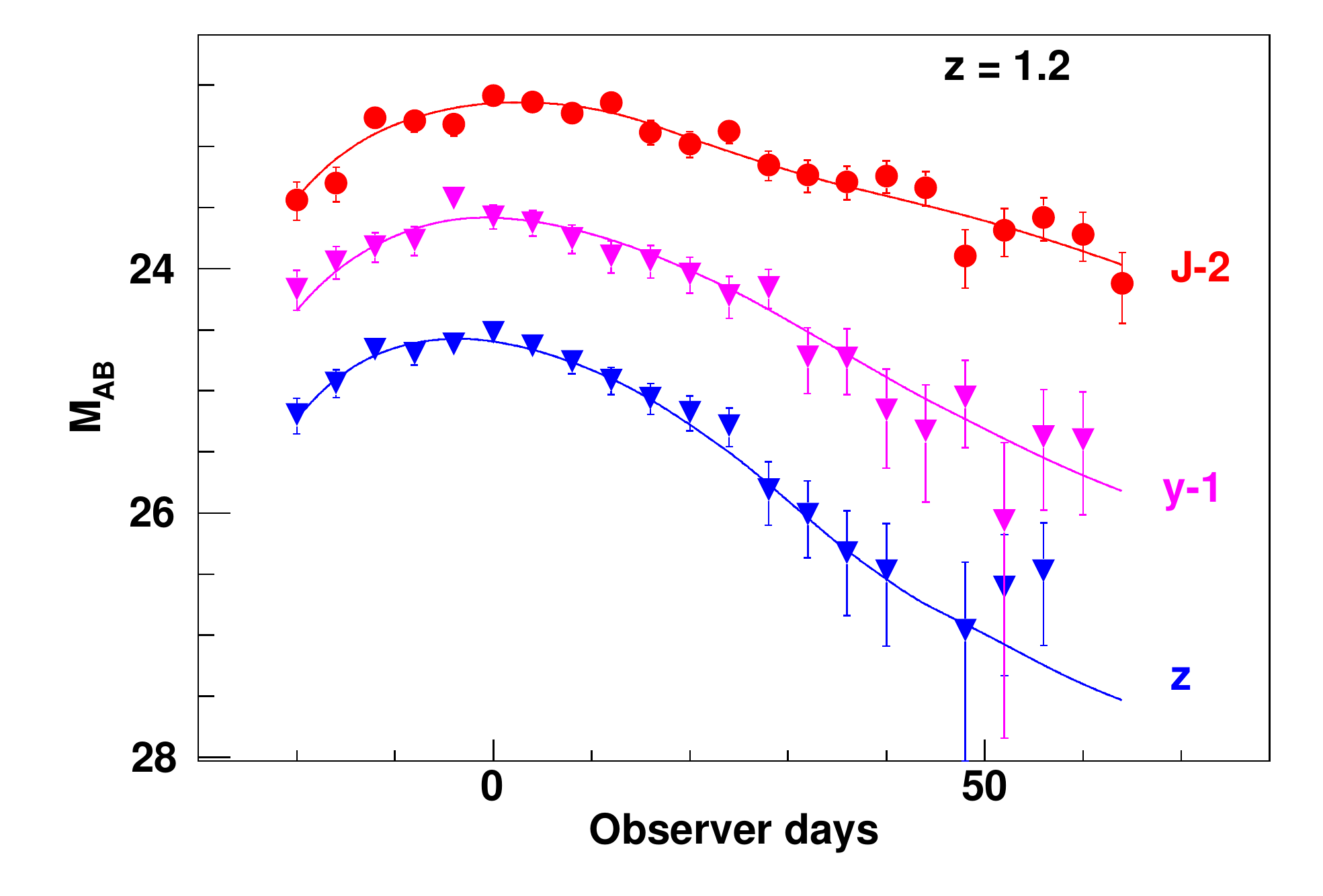}
\includegraphics[width=\linewidth]{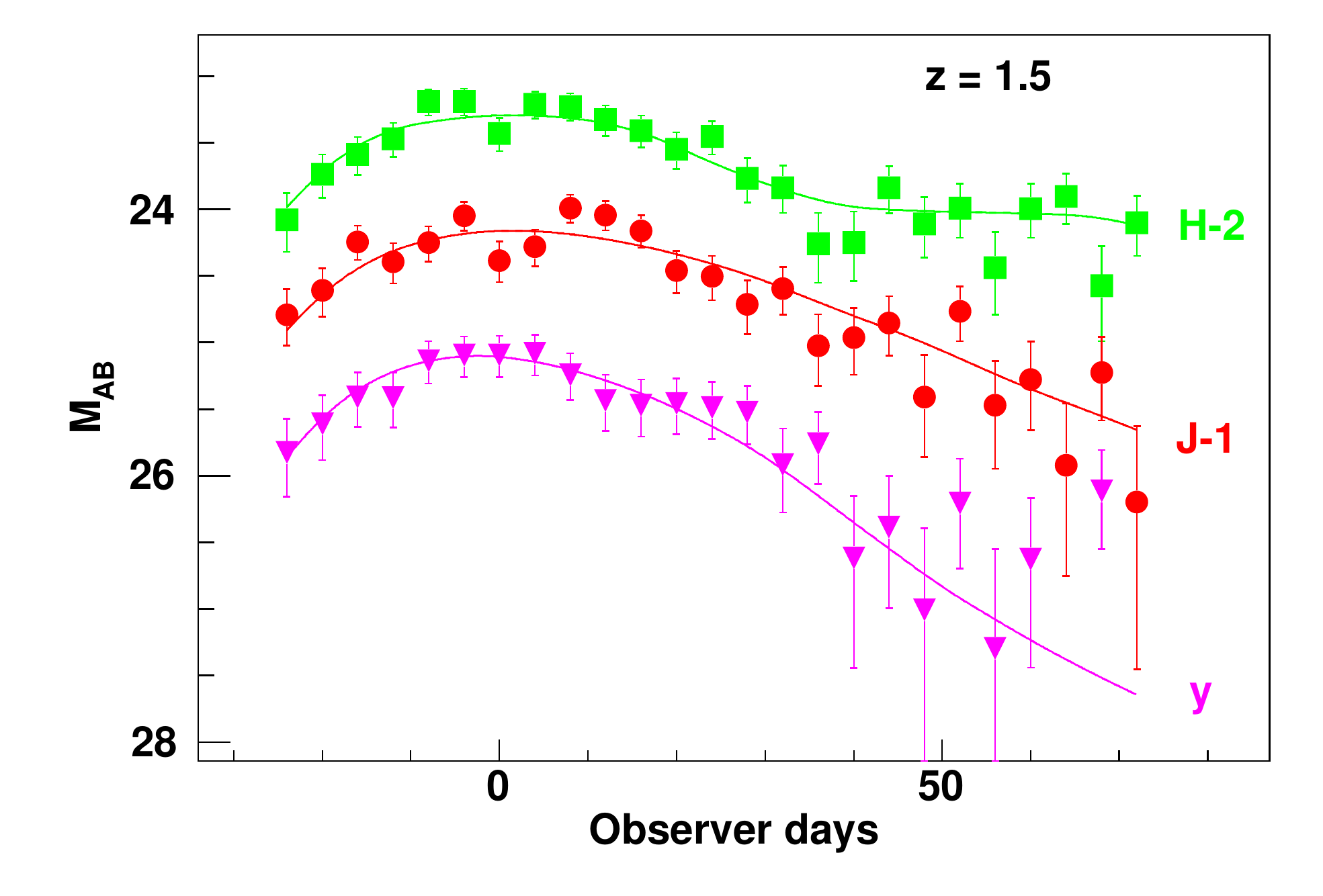}
\caption{Simulated light curves of an average SN at $z=1.2$ (top) and $z=1.5$ (bottom).
\label{fig:high-z-light curves}}
\end{figure}

In Table~\ref{tab:joint-exp-times}, we display the depth per visit
that delivers the required quality of light curves up to $z=1.5$ (for an
average SN). This table also lists observing times derived using our
instrument simulator. For $i$ and $z$ band, we used the sensitivities 
used for LSST simulations from \cite{LivingLSST}, however without accounting
for the IQ degradation with air mass: somewhat longer exposure times
might be needed in order to reach the required sensitivities. 
As for the Euclid observations, a slower cadence could be accommodated provided the depth per visit is increased accordingly.
The derived precision of single-band light curve amplitudes of average
SNe~Ia are displayed in Fig.~\ref{fig:jointsurvey_reso}. Examples
of simulated light curves are shown in Fig.~\ref{fig:high-z-light curves}.
\begin{table}[h]
\caption{Depth of the visits simulated for the DESIRE survey.}
\begin{center}
\begin{tabular}{|l|ccccc|}
\hline
  & i & z  & y & J & H \\
\hline
Depth (5$\sigma$) & 26.05 & 25.64 & 25.51 & 25.83 & 26.08 \\
Exp. time (s) & 700 & 1000 & 1200 & 2100 & 2100 \\
\hline
\end{tabular}
\tablefoot{Depth (5$\sigma$ for a point source) 
and exposure times at each visit for a 4-day cadence of the
proposed DESIRE joint SN survey. The exposure times 
for LSST $i$ and $z$ bands assume nominal observing conditions. For Euclid NIR bands,
the exposures times are the ones that would deliver the required depth
in a single exposure, if such long exposures are technically possible.
The S/N calculations are described in appendix \ref{sec:point-source-photometry}.
\label{tab:joint-exp-times}}
\end{center}
\end{table}

We have assumed that Euclid could devote 6 months of its programme to monitor
this dedicated deep field, possibly within an extended mission. The NIR exposure times in
Table~\ref{tab:joint-exp-times} add up to 5400 s per visit and
pointing.  Monitoring 20 deg$^2$ (40 pointings) at a four-day cadence
uses 62.5\% of the wall clock time for integrating on the sky. 
The rest is available for overheads such as readout, slewing, etc.
Since
building SN light curves require images without the SN, the programme
is split over two seasons with identical pointings, so that each 
season, which consists of 45 visits, provides a deep SN-free image for the other season.
Thus, our baseline programme consists of two
six-month seasons, where the SN survey is allocated half of the clock
time. In practice, this means that the same 10 deg$^2$ field will be observed
twice, in two 6-month seasons during which the field should be visible from the ground. Within this scheme, the reference images (i.e. images without
the SN) gather on average 1.5 observing season (i.e. 67 epochs for a
4-day cadence). We accounted for the finite reference depth effect of
Euclid images assuming a 60-epoch reference (i.e. 1.3 season),
following the algebra provided in appendix~\ref{app:ref-depth}.
Regarding reference depth, the situation for ground-based surveys is
different since those are planning 5 \citep[for the DES SN survey,
  see][]{Bernstein12} to 10 \citep[for LSST, see][]{LivingLSST}
observing seasons on the same field. The effect then amounts to a less
than 10\% degradation of amplitude measurements due to shot noise,
which is sub-dominant in most of the redshift range, and we neglected
the effect.

\begin{center}
\begin{figure}[h]
\includegraphics[width=\linewidth]{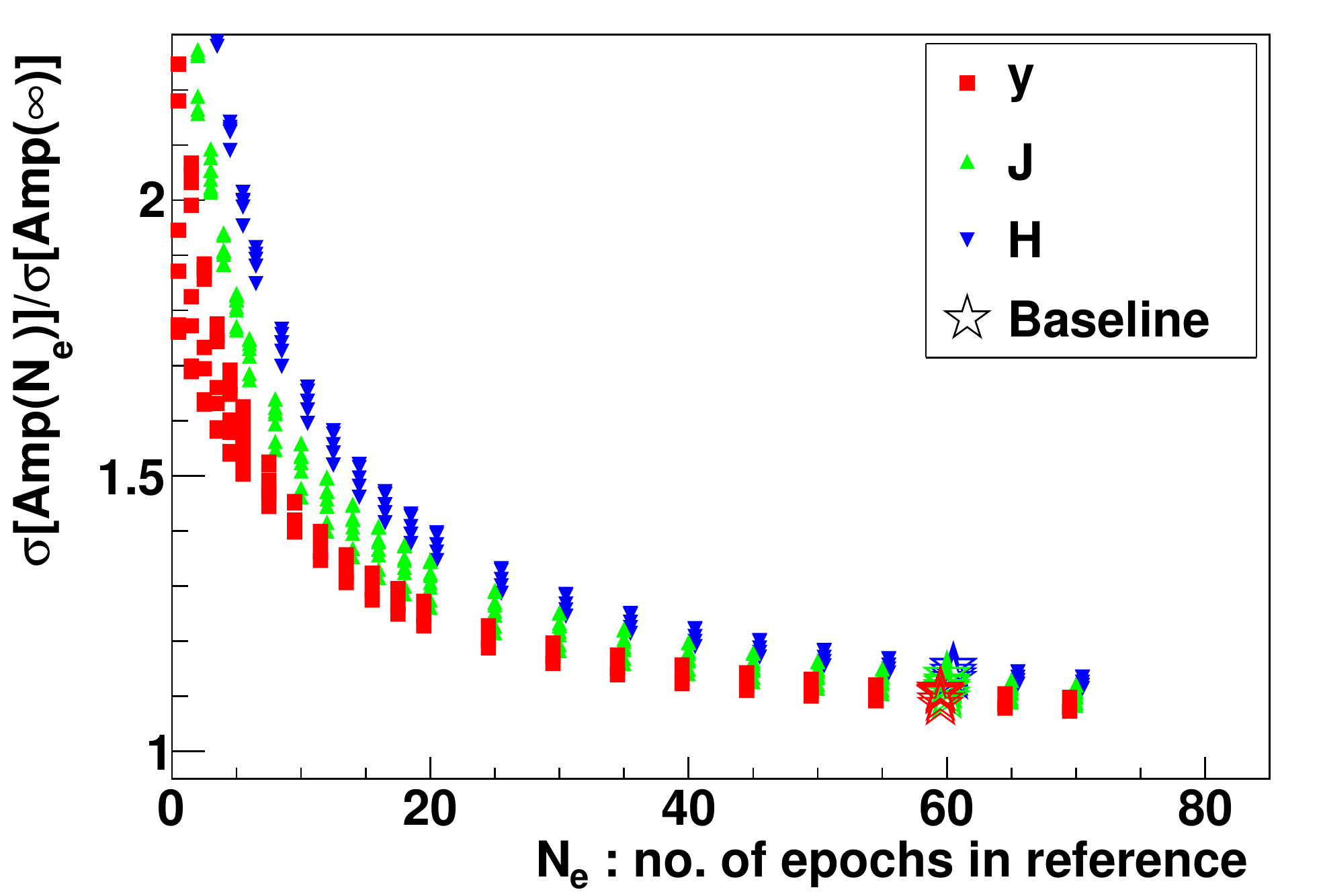}
\caption{Precision of light curve amplitude measurement, in units of
  the measurement quality for an infinitely deep reference, as a
  function of the number of epochs $N_e$ used in the reference image.
  For each band, the spread at a given reference depth is due  
  to redshift ($0.75<z<1.55$), and the effect increases with redshift.
  If all events were measured using 45 reference epochs (i.e. one season),
  the measurement precision would degrade by less than 10 \% relative to the
  chosen baseline, i.e. 60.
\label{fig:PlotRefEpochs}}
\end{figure}
\end{center}

Regarding light curves in Euclid bands, we varied the reference depth
in order to assess the acceptable variations of this parameter, and
we display the impact of different reference depths in
Fig.~\ref{fig:PlotRefEpochs}.  Beyond 45 epochs (i.e. one season),
the actual number does not make a large difference with our baseline.
On the contrary, scenarios with a reference shallower than 15 to 20 epochs 
seriously degrade the measurement quality.

It is mandatory that the chosen field is observable by both Euclid and
a ground-based observatory. The former imposes a field close to the
ecliptic poles. The southern ecliptic pole suffers from Milky Way
extinction and a high stellar density, but there are acceptable
locations within 10$^\circ$ from the pole, observable for 6 months or
more from the LSST site. The amount of observing time for LSST is
modest, and could even be included as one of its ``deep-drilling
fields'', which are already part of its observing plan. DECam on the
CTIO Blanco could likely deliver the required sky coverage and depth
in less than a night every 4th night. The northern ecliptic pole is
observable by the Subaru telescope.

\subsection{Other SN surveys by the time Euclid flies}

By the time Euclid flies, we expect that the Dark Energy Survey (DES)
will have produced a few thousand supernovae extending to $z\sim 1$
\citep{Bernstein12}.  LSST is not constructed yet, but it is expected
to be a massive producer of SN light curves in the visible. LSST can
tackle two redshifts regimes. First is the $0.2\lesssim z \lesssim 1$ regime already covered by
ESSENCE \citep{WoodVasey07}, SNLS \citep{Sullivan11}, and Pan-STARSS \citep{Scolnic14, Rest14}, and by DES in the near future. Second
is 
the ``nearby'' redshift regime, where LSST's large étendue and fast readout
allow it to rapidly cover large areas of sky. We now sketch 
a plausible contribution of LSST to the Hubble diagram of SNe~Ia
in these two redshift regimes.

\subsubsection{LSST Deep-Drilling Fields}
\begin{center}
\begin{figure}[h]
\includegraphics[width=\linewidth]{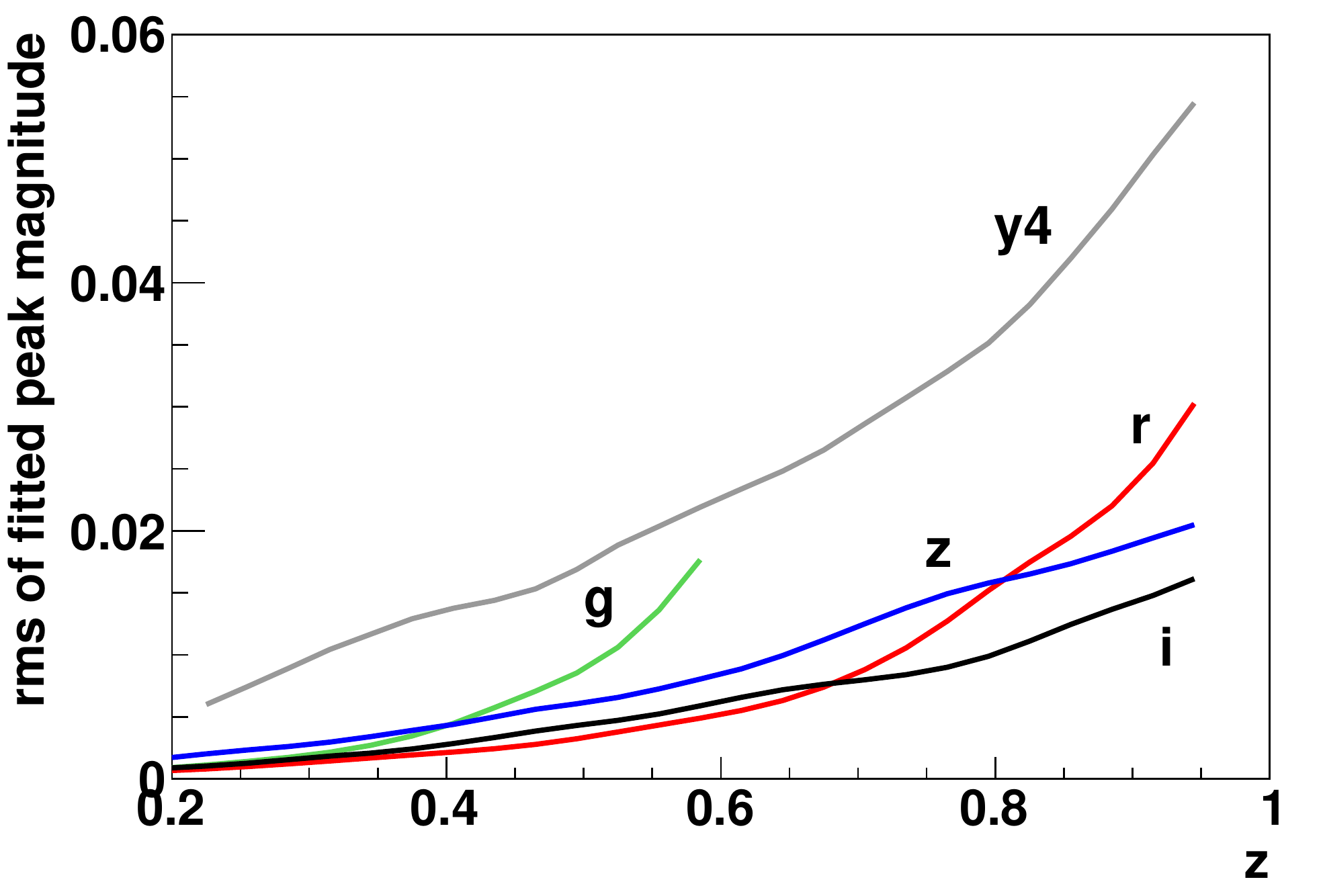}
\caption{Precision of light curve amplitudes as a function of redshift
  for the 5 bands of the LSST deep-drilling-fields survey, assuming a 4-day cadence
  with the depths from Table~\ref{tab:lsstddf-times}. At the
  anticipated depth, the contribution of the $y4$ band is marginal for
  distances to SNe. It however provides us with 3 bands within
  requirements at the highest redshift.
\label{fig:lsstddf_reso}}
\end{figure}
\end{center}

The LSST deep-drilling fields (DDF) observations cover several scientific
objectives, including distances to SNe. The current baseline for the
observations consists of an approximately 4-day cadence with
exposure times provided in Table~\ref{tab:lsstddf-times}. The
corresponding fitted amplitude precisions are displayed in 
Fig.~\ref{fig:lsstddf_reso}.  The limiting redshift for a three-band
measurement above 380 nm (restframe) is $z\simeq 0.95$, where the quality of
$r$-band is more than adequate for identification. We note that the
precisions displayed in Fig.~\ref{fig:lsstddf_reso} leave a good
margin for less-than-optimal observations: a moderate degradation
of image quality or time sampling would not affect our conclusions.

\begin{table}[h]
\begin{center}
\caption{Simulated depths per visit of the LSST Deep Drilling Fields}
\begin{tabular}{|l|ccccc|}
\hline
                  & g     & r     & i      & z     & y4 \\
\hline
depth (5$\sigma$) & 26.47 & 26.35 & 25.96  & 25.50 & 24.51 \\
Exp. time (s)     & 300   & 600   & 600    & 780   & 600 \\
\hline
\end{tabular}
\tablefoot{
  The exposure times refer to dark and otherwise 
  average observing
  conditions.  The $y4$ filter is the widest considered option for the 
LSST $y$-band.
\label{tab:lsstddf-times}}
\end{center}
\end{table}

The volume of LSST deep-drilling fields  observations adequate 
for distances to SNe is
not settled yet but the current goal consists of monitoring 4 fields
for 10 seasons. We conservatively assumed the statistics corresponding
to 4 fields (each of 10 deg$^2$) monitored over five 6-months
seasons. 5 fields over 4 seasons yield the same event statistics.

\subsubsection{Low-redshift supernovae with LSST}
Cosmological constraints from relative distances enormously benefit
from a local measurement and essentially all cosmological constraints
from the Hubble diagram of SNe Ia make use of a nearby SN
sample. Since the relative calibration between surveys is currently a
serious limitation \citep[see e.g.][]{Conley11}, we might wonder whether
LSST itself might collect such a nearby sample. The LSST wide survey
is built from two 15s exposure visits \citep{LivingLSST}, and covers 20,000 deg$^2$. The depth required to
measure the shear field and photometric redshifts for galaxies is
eventually obtained from several hundred exposures. If these exposures
are evenly spread over 10 years, the time sampling is too coarse 
to measure distances to SNe~Ia. We argue here that an uneven time sampling would 
allow us to monitor some fields with a 4-day cadence within the same
overall time allocation (and hence final depth): one or more seasons
observed at a $\sim$ 4-day cadence using the regular LSST observing
block (2$\times$15 s) would deliver a depth per visit slightly
higher than the SDSS-II SN survey \citep{Kessler09,Sako14}. Since LSST aims at monitoring
20,000 deg$^2$ for 10 years, we conservatively assumed that a proper
cadence for SNe might be acquired over 3000 deg$^2$ for 6 months,
which amounts to $\sim$~10 times the volume of the SDSS-II SN survey. We only consider
events at $0.05<z<0.35$ where the quality is safely within
requirements of \S\ref{sec:lc-requirements}. The lower redshift bound eliminates worries about
peculiar velocities significantly affecting redshifts.
The upper redshift bound derives from the cadence we have assumed and 
the depth of LSST visits. 
We note that this kind of observing strategy is not adopted yet
within LSST, although it is actively studied. 
It might be implemented because it allows for
additional science that cannot be done with evenly distributed sampling,
and with no additional observing time.

The imposed quality requirements imply that all surveys 
are able to detect many events beyond their assigned high-redshift 
cutoff. This allows us to work in the
redshift-limited regime in order to capture a similar
fraction of the SN population at all redshifts.
To summarise, we provide the main parameters for the three surveys 
in Table~\ref{tab:summary-surveys}.
\begin{table}[h]
\caption{Main parameters of the simulated surveys.}
\begin{center}
\begin{tabular}{|l|rrrrr|}
\hline
  & $z_{min}$ & $z_{max}$  & area & duration& events \\
  &          &           & (deg$^2$) & (months)  &  \\
\hline
DESIRE & 0.75 & 1.55 & 10 & 2x6 & 1740 \\
LSST-DDF           & 0.15 & 0.95 & 50 & 4x6  &  8800 \\
Low z           & 0.05 & 0.35  & 3000 & 6 & 8000 \\
\hline
\end{tabular}
\tablefoot{The duration of the DESIRE survey
is two times 6 months, but the Euclid observations use only half of the 
clock time, and so add up to 6 months of clock time.
\label{tab:summary-surveys}}
\end{center}
\end{table}

\section{Redshifts and SN classification\label{sec:redshifts-typing}}

\subsection{Redshifts\label{sec:redshifts}}
With the statistics we are considering, we cannot expect to classify
spectroscopically all events entering the Hubble diagram, as most of
the SN surveys have done up to now. Spectroscopy remains, however, the
only way to acquire an accurate redshift, and we will assume in what
follows that host galaxy spectroscopic redshifts are acquired at some
point, possibly after the fact, using multi-object spectroscopy. The
4MOST and DESI projects on 4~m telescopes would both be well suited to
obtaining spectroscopic redshifts of the majority of the host
galaxies, as demonstrated by the sucessful use of the AAOmega
instrument on AAT to observe host galaxies from SNLS \cite{Lidman13}.
Host galaxies remaining with unmeasured redshifts after such a
campaign would be followed up with optical and infrared spectroscopy
on 8m or Extremely Large Telescopes.

In order to evaluate the required exposure times to acquire
  host redshift with a multi-object spectrograph, and the efficiency
  at obtaining host redshifts in SN surveys, we have studied how
  spectroscopic redshifts were assigned to a subsample of the SNLS
  events. We have selected SNLS spectra to $0.5<z<1$, which can be 
  ``translated'' to $0.75<z<1.55$ by multiplying luminosity distances by
  1.65, in order to emulate collection of host redshifts in the DESIRE 
  survey. We have examined 40 slit spectra of ``live'' SNe collected
  using FORS2 on the VLT, and the origin of redshift determination splits this
  ``training'' sample into three event classes:
\begin{itemize}
\item[-] 20 events happened in emission line galaxies (ELGs) where the
  redshift was obtained from the [\ion{O}{ii}] doublet (3726 \& 3729 \AA, unresolved with
  FORS2). We have then measured the  [\ion{O}{ii}] line intensity.
\item[-] 11 events happened in passive hosts and the redshift was
  obtained from the Ca H\& K absorption lines (3933 \& 3968 \AA). In these cases, we
  collected the host magnitudes from imaging data.
\item[-] 9 events did not have enough galaxy flux in the slit and
  were assigned a redshift using supernova features.
\end{itemize}

We note that both the  [\ion{O}{ii}] doublet and the Ca H\&K lines remain within
the wavelength reach of (deep-depleted) silicon sensors at $z=1.55$. In
order to derive exposure times at higher redsdhifts than our SNLS subsample, 
we rely on the BigBoss (now called DESI) proposal \citep{BigBossProposal09}. Namely, this proposal
evaluates that with a 1000-s exposure time, it is possible to detect
each member of the  [\ion{O}{ii}] doublet at a S/N of 8 if the  [\ion{O}{ii}] brightness is
0.9 $10^{-16}$ergs/s, at a fairly extreme redshift of 1.75. Drawing
from the SDSS experience, the same proposal relates passive supernova
magnitudes and the exposure time required to get a redshift.  We have
``translated'' the  [\ion{O}{ii}] line brightnesses and host magnitudes of our
test sample to higher redshifts by increasing the luminosity distance
by 1.65, and using the BigBoss figures, we have evaluated that DESIRE
emission line host galaxies would require up to 300 ks to deliver
S/N=8 per  [\ion{O}{ii}] doublet member, and passive hosts would require up to
100 ks to deliver a redshift. Most of the hosts would require
significantly less. These derived exposure times compare well with the
extrapolation of the typical $\sim$3600-s on FORS2 of the SNLS
spectra: this exposure time translates to $\sim$ 100 ks for DESIRE
host spectroscopy on DESI when accounting for the mirror size ratio
($\sim 2^{2}$ for the diameter) and the fainter targets ($\sim 3^{2}$). We find that the S/N
requirement of the DESI proposal for  [\ion{O}{ii}] emitters is higher than the ones 
we obtained for the faintest members of our training sample.

These exposure times might look large, but one should note that in
\cite{Lidman13}, exposure times of 90 ks are reported. The redshift
reach of the \cite{Lidman13} pioneering programme does not extend
significantly at $z>1$, because it targeted hosts of SN candidates
detected in the SNLS imaging data \citep{Bazin11}, limited in redshift
by the poor red sensitivity of the Megacam sensors (see
e.g. \citealt{MegacamPaper}). One might also note that the ultra deep
VIMOS survey (50 ks exposures on the VLT) obtained a success rate at
obtaining redshifts \citep{LeFevre14} similar to our anticipation.
Regarding the  [\ion{O}{ii}] line brightness of SN
hosts, three features indicate that our estimation is conservative;
first, the SNLS spectroscopic campaign aimed at identifying live
supernovae and the slit position was firstly aimed at maximising the
SN flux, with less consideration for the host. Our training  [\ion{O}{ii}]
luminosities are then likely to be underestimated, as compared to
fibre-fed spectroscopy targeting the host galaxy; second, the average
 [\ion{O}{ii}] brightness of ELGs tend to increase with z, and SN hosts likely
follow this trend; third, as already mentioned, we are able to measure
host redshifts at S/N lower than 8 per  [\ion{O}{ii}] doublet member. So, we
estimate that typically 75\% of
DESIRE host redshifts could be secured by means of multi-fibre
spectroscopy in the visible. Fainter hosts could be targeted by more
powerful instruments, and spectra of a subsample of the supernovae
themselves (\S\ref{sec:sn-spectra}) almost unavoidably deliver
redshifts.

One might consider the possibility of relying on photometric redshifts
of supernovae. These are now known to be significantly more accurate
than photometric redshifts of host galaxies
\citep{Palanque10,Kessler10}, thanks to the homogeneity of the
events. SN photometric redshifts however introduce correlated
uncertainties between distance and redshift which would require a
careful study. SN photometric redshifts also degrade the performance
of photometric identification and classification.

\subsection{SN spectra \label{sec:sn-spectra}}
Spectra have been used to obtain detailed information on supernovae,
mostly to empirically compare high- and low-redshift spectra (e.g.
\citealt{Maguire12} and references therein). We can consider extending 
these comparisons to higher redshifts, relying on future
facilities: both ground-based extremely large telescopes and
the JWST will allow one to efficiently acquire NIR good-quality 
spectra of SNe~Ia at $z\sim1.5$ \citep{Hook12}. 
Using the available Exposure Time Calculators,
we have evaluated exposure times of 900s for the E-ELT, and 1500s
for prism spectroscopy using NIRSPEC on JWST to acquire a spectrum
of an average SN~Ia at $z=1.55$, with a quality sufficient
to compare spectral features with lower redshifts. We anticipate 
similar integration times with the 23-m Giant Magellan Telescope 
and the (30-m) Thirty Meter Telescope. These integration times 
are significantly lower than the typical 2 hours required to identify 
a $z\sim 1$ SN~Ia event using an 8-m class (ground-based) telescope.

In current surveys, SN spectra are primarily used to identify the events
(see e.g. \citealt{Howell05,Zheng08}). Although we cannot hope to reproduce
this strategy, obtaining SN spectra of a subsample will help
characterise the transient population and in particular the
interlopers of the Hubble diagram. Given the exposure times
above, assuming 40 h per semester awarded on both an ELT and JWST,
and typically 30 mn per target including overheads, 
we could collect typically 300 live SN spectra. For the brightest targets,
large programmes on existing 8-10~m telescopes could deliver $\sim$ 200
spectra if 400 hours could be gathered in total. So, collecting several hundred
spectra of DESIRE events is a plausible goal.

\subsection{SN classification}
Most core-collapse supernovae are fainter than SNe Ia, and exhibit a
larger luminosity dispersion. In \S~3.5 of A11, following arguments
developed in \cite{Conley11}, it is shown that iteratively clipping to
$\pm 3 \sigma$ the contaminated Hubble diagram yields acceptable
biases to the distance redshift-relation, under various contamination
hypotheses. This crude approach works because the contamination
contribution to the Hubble diagram does not evolve rapidly with
redshift. This crude typing conservatively assumes that
light curve shapes and colours do not provide type information. Although all recent
SN analyses indeed clip their Hubble diagram, we regard this
purification through clipping as a backup plan, and we would prefer a
selection based on colours and light curve shapes as proposed in
e.g. \cite{Bazin11,Sako11,Campbell13}. The high photometric quality
requirements we are imposing are an obvious help in this respect.
Any method used to purify the Hubble diagram sample will be cross-checked
using spectra of a subsample of active SNe, see \S~\ref{sec:spectro-UV}.

\section{Forecast method \label{sec:methodology}}
In order to derive cosmological constraints, we follow the methods developed
for A11, with the aim of accounting as precisely as possible for
systematic uncertainties, including the interplay between different
uncertainty sources. We will discuss astrophysical issues associated
with SNe~Ia distances in \S~\ref{sec:astro-issues}, and discuss here uncertainties
mostly associated with the measurements themselves. In our forecast, we account for photometric calibration
uncertainties, statistical light curve model uncertainties (because
the training sample is finite), and photometric calibration
uncertainty of the training sample, residual scatter around the model,
fit of the brighter-slower and brighter-bluer relations, and make some
provision for irreducible distance errors. We account
for systematic uncertainties using nuisance parameters,
and build a
Fisher matrix for all parameters (including SN event parameters) that
we invert in order to extract the covariance of the cosmological
parameters. This gives us cosmological uncertainties marginalised over
all other parameters. The method is detailed in A11 and we list now
the considered uncertainty sources (and their size when applicable):
\begin{itemize}
\item The measurement shot noise.
\item Statistical uncertainties of the light curve model. We assumed that
it is trained on the cosmological data set. 
\item Systematic uncertainties due to flux calibration, both on SN
  parameters and through the SN model training. Our baseline assumes that
  the conversion of measured counts to physical fluxes is uncertain at
  the 0.01 mag level rms, independently for each band in visible and NIR. This level is
  conservative for the visible range considering the 
  accuracy reached in \cite{Betoule13}. The impact of varying the photometric calibration accuracy is discussed in \S~\ref{sec:alteration}.
\item The intrinsic scatter of supernovae at fixed colour (called
  colour smearing). We assign (magnitude) rms fluctuations of
  broadband amplitudes of 0.025, following Fig.~8 of \cite{Guy10}.
  We note that a more optimistic value $\sigma_c=0.01$ was assumed in
  \cite{Kim06}.  Larger smearings are indeed observed in the UV, but we
  ignore bands with $\bar{\lambda}<380$~nm (where $\bar{\lambda}$ is
  the central wavelength of the filter).
\item We fit for both brighter-slower and brighter-bluer relations
and marginalise over their coefficients.
\item We assume an intrinsic distance scatter of 0.12 mag, where
current estimates are around or below 0.10 \citep{Guy10}. The average 
  Hubble diagram residual is about 0.14 rms, where
  the difference to 0.12 is mainly due to colour smearing. 
  
\item We assume that there is an irreducible distance modulus error,
affecting all events coherently, varying 
linearly with redshift,
\begin{equation}
\delta \mu = e_M \times z, \label{eq:em_times_z}
\end{equation}

with a Gaussian prior $\sigma(e_M)$=0.01. This distance modulus error
accounts for possible evolution of SNe~Ia with redshift, not accounted
for by the distance estimator, which in turn biases the measured
distance-redshift relation. In A11 (\S 5.2), a metallicity indicator
relying on UV flux is proposed that allows one to control the distance
indicator at the level of $\sim 0.01$. Our ansatz above (Eq.~\ref{eq:em_times_z}) makes provision
for $\delta \mu = 0.015$ over the whole redshift range.

\end{itemize}

These uncertainties describe current know-how, in a rather
conservative way. It is thus likely that we might eventually do
better. The code that implements the global fit successfully
reproduces the SNLS3 uncertainties.

In order to propagate uncertainties, we introduce nuisance
parameters in the fit (e.g. alteration to the photometric zero points) and
eventually marginalise over those.  In order to emulate the light curve fitter
training and the impact of calibration uncertainties, event parameters
are also fitted, together with offsets to the fiducial SN
model. Appendix~A of A11 compares the propagation of uncertainties
and the introduction of nuisance parameters and concludes that both
approaches are strictly equivalent. Our global fit thus considers 5 sets of
parameters:
\begin{itemize}
\item The event parameters in their SALT2 flavour:  $t_0$ is a reference date, $X_0$ 
is the overall brightness, $X_1$ indexes light curve shape, and $c$ is a rest-frame
  colour. 
\item The photometric zero points, or more precisely offsets to
  their nominal values. We impose priors on these offsets which
  account for photometric calibration accuracy,
  from SN instrumental fluxes to physical fluxes.
\item The global parameters $(\alpha,\beta, {\mathcal M})$ 
used to derive a distance from the SN
  parameters:
\begin{equation}
\mu = m^*_B+\alpha X_1 - \beta c - {\mathcal M},
\label{eq:mu-tripp}
\end{equation}
Following Eq.~\ref{eq:em_times_z}, we emulate an irreducible fully correlated distance error with
\begin{equation}
{\mathcal M}={\mathcal M}_0 + e_M \times z,  \label{eq:M_depend_z}
\end{equation}
where $e_M$ is constrained
with a Gaussian prior of rms 0.01. The actual parameters are hence
$(\alpha,\beta,{\mathcal M}_0, e_M)$. 
The overall flux scale of the Hubble diagram is
unknown and ${\mathcal M_0}$, which is marginalised over, accounts for it.
\item The supernova model definition. We model both the peak brightness
of the average SN as a function of wavelength, and how colour variations affect different wavelengths (see \S~4.3.3 of A11). 
For both quantities we model offsets to the fiducial SN model, 
using 10-parameter polynomials over the SN model restframe
spectral range, which makes more than 2 parameters per regular
  broadband filter. These parameters account for the SN model training.
\item The cosmological parameters.
\end{itemize}

SN cosmology usually proceeds in two steps: first extracting event
parameters by fitting a model to light curves, and then fitting
cosmology to distances derived from these event parameters.  With
calibration uncertainties at play, the first step results in fully
correlated event parameters, but the correlations due to systematics
are in fact a small-rank matrix, compared to the number of events we
are considering here.  Instead of this two-step procedure, we carry
out both steps simultaneously summing all terms in a single
$\chi^2$, where the light curve fit term also incorporates the light curve fitter
training. This method is equivalent to the two-step procedure but
does not require propagating a large covariance or weight 
matrix between stages.
The fit involves a large number of parameters (more than 50 000)
and the appendix B of A11 sketches the method used to compute
the covariance matrix of a small subset of parameters, among which
are the cosmological parameters.

\section{Forecast results\label{sec:results}}
In order to evaluate the cosmological constraints that the proposed surveys
could deliver, we use the commonly used equation of state (EoS)
effective parametrisation proposed in \cite{ChevallierPolarski01}:
$w(z) = w_0 + w_a\ z/(1+z)$, and shown to describe a wide array of
dark energy models in \cite{Linder03}. We define the cosmology with two more
parameters: $\Omega_M$ the reduced matter density, and $\Omega_X$ the
reduced dark energy density, both evaluated today. Distances alone do not constrain
efficiently these 4 parameters, and in practice, at least two external
constraints have to be added. We have settled for one CMB prior, taken
as a measurement of the shift parameter $R \equiv \om^{1/2}
  H_0 r(z_{CMB})$, and flatness.  For the geometrical CMB prior,
we compared the $R$ measurement to 0.32\% (anticipated from {\it Planck}, 
see \citealt{Mukherjee08},
Table~1), with the binned $w$ matrix for CMB alone from \cite{JDEMFoM09}
projected on the $(w_0,w_a)$ plane in a flat universe, and found
extremely similar results. Both approaches take care to
ignore information on dark energy from the ISW effect in the CMB,
because the latter concentrates on large angular scales and might be
difficult to extract. We also wish to ignore the ISW effect in
order to ensure a purely geometrical cosmological measurement
that is insensitive to the growth of structures after decoupling.
The method also ignores potential information from CMB lensing.
We describe in appendix \ref{sec:binned-matrix} how
to obtain SN-only constraints from our results.

We simulate distances in a fiducial flat ${\rm \Lambda}$CDM universe
with $\Omega_{\rm M}=0.27$. We restrict the
rest frame central wavelength of the bands entering the fit to
[380-700]nm, which leaves 3 to 4 bands per event.
Enlarging this restframe spectral range formally improves the
statistical performance but breaks the requirement that similar rest frame ranges
are used to derive distances at all redshifts.

\begin{figure}[h]
\centering
\includegraphics[width=\linewidth]{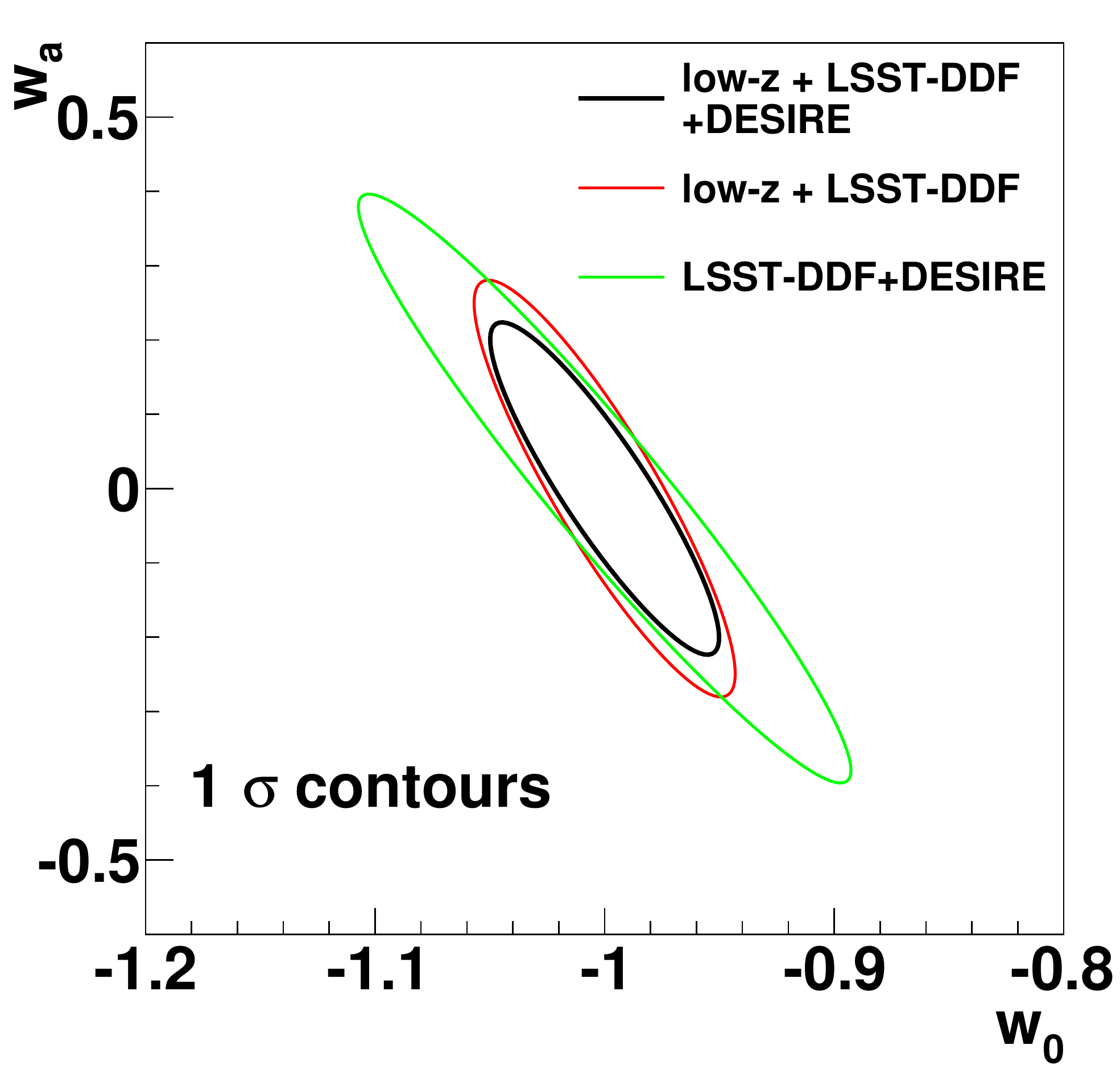}
\caption{Confidence contours (at the 1$\sigma$ level) of the survey 
combinations listed in Table~\ref{tab:cosmo-fom}. 
The assumptions for systematics correspond to the last row of Table~\ref{tab:aaa-sys-combinations}.
\label{fig:ellipses}
}
\end{figure}

The quality of EoS constraints are usually expressed, following
\cite{DETF06}, from the area of the confidence contours in the
$(w_0,w_a)$ plane, and the normalisation we adopt reads
$FoM=[Det(Cov(w_0, w_a))]^{-1/2}$.  Still following \cite{DETF06}, we
define the pivot redshift $z_p$ to be where the EoS uncertainty
is minimal, and $w_p \equiv w(z_p)$. $\sigma(w_p)$ is also the
uncertainty when fitting a constant EoS.  $\sigma(w_p)$ can be
regarded as the ability of the proposed strategy to challenge the
cosmological constant paradigm. In Table~\ref{tab:cosmo-fom}, we
report the following performance indicators: $\sigma(w_p)$, the uncertainty of
the EoS evolution $\sigma(w_a)$, and the FoM. The FoM difference
between the two first lines shows the Euclid contribution to the overall
FoM: by delivering about 10\% of the total event statistics (see Table~\ref{tab:summary-surveys}), the high
redshift Euclid part of the Hubble diagram increases the FoM by
$\sim$50\%. The confidence contours corresponding to 
 Table~\ref{tab:cosmo-fom} rows are displayed in 
Fig. \ref{fig:ellipses}.

\begin{table}[h]
\caption{Cosmological performance of the simulated surveys.}
\begin{center}
\begin{tabular}{l|rrrr|}
  &  {\boldmath $\sigma(w_a)$} &  {\boldmath $z_p$} &  {\boldmath$\sigma(w_p)$} & {\bf FoM} \\ 
\hline
low-z + LSST-DDF  & \multirow{2}{*}{0.22} & \multirow{2}{*}{0.25} & \multirow{2}{*}{0.022} & \multirow{2}{*}{203.2} \\  
        \multicolumn{1}{c|}{+ DESIRE} & & & & \\
low-z + LSST-DDF &  0.28 &  0.22 &    0.026 &   137.1 \\
LSST-DDF + DESIRE &   0.40 &  0.35 &    0.031 &    81.4 \\
\hline
\end{tabular}
\tablefoot{The FoMs assume a 1-D geometrical {\it Planck} prior and flatness.
$z_p$ is the redshift at which the equation of state uncertainty
reaches its minimum $\sigma(w_p)$. The FoM is defined as $[Det(Cov(w_0, w_a))]^{-1/2}= [\sigma(w_a)\sigma(w_p)]^{-1}$ and accounts for systematic uncertainties.
The contributions of the main systematics are detailed in 
Table~\ref{tab:aaa-sys-combinations}.
\label{tab:cosmo-fom}}
\end{center}
\end{table}

\begin{table}[h]
\caption{Cosmological performance with various uncertainty sources.}
\begin{center}
\begin{tabular}{ccc|cccc|}

\multicolumn{3}{c|}{\bf Assumptions} &  \\
{\bf cal} & {\bf evo} & {\bf train} &    {\boldmath $\sigma(w_a)$} &  {\boldmath $z_p$} &  {\boldmath$\sigma(w_p)$} & {\bf FoM} \\ 
\hline
n & n & n & 0.15 & 0.30 & 0.016 & 418 \\
y & n & n & 0.18 & 0.30 & 0.016 & 339 \\
n & y & n & 0.18 & 0.25 & 0.018 & 315 \\
y & y & n & 0.20 & 0.27 & 0.019 & 266 \\
n & n & y & 0.16 & 0.30 & 0.016 & 403 \\
n & y & y & 0.18 & 0.25 & 0.018 & 304 \\
y & n & y & 0.21 & 0.28 & 0.020 & 238 \\
y & y & y & 0.22 & 0.25 & 0.022 & 203 \\
\hline
\end{tabular}
\end{center}
\tablefoot{``cal'' refers to calibration uncertainties ($\sigma_{\rm ZP}=0.01$).
``evo'' refers to evolution systematics (Eq.~\ref{eq:em_times_z}).
``train'' refers to SN model training from the same sample.
 \label{tab:aaa-sys-combinations}}
\end{table}

We present in Table~\ref{tab:aaa-sys-combinations}
some combinations of uncertainties, and we find (as in Table~5 of A11)
that the dominant reduction in the figure of merit arises from
the combination of calibration uncertainties and SN model training. In A11,
we also considered the impact of several hypotheses such as fitting
the $\alpha$ and $\beta$ parameters (equation \ref{eq:mu-tripp})
separately in redshift slices, or assuming that there are several
event species, each with its light curve model and ($\alpha,\beta,
{\mathcal M_0, e_M}$) set, and concluded that these extra parameters result
in negligible degradation of the cosmological precision.

\begin{figure}[h]
\centering
\includegraphics[width=\linewidth]{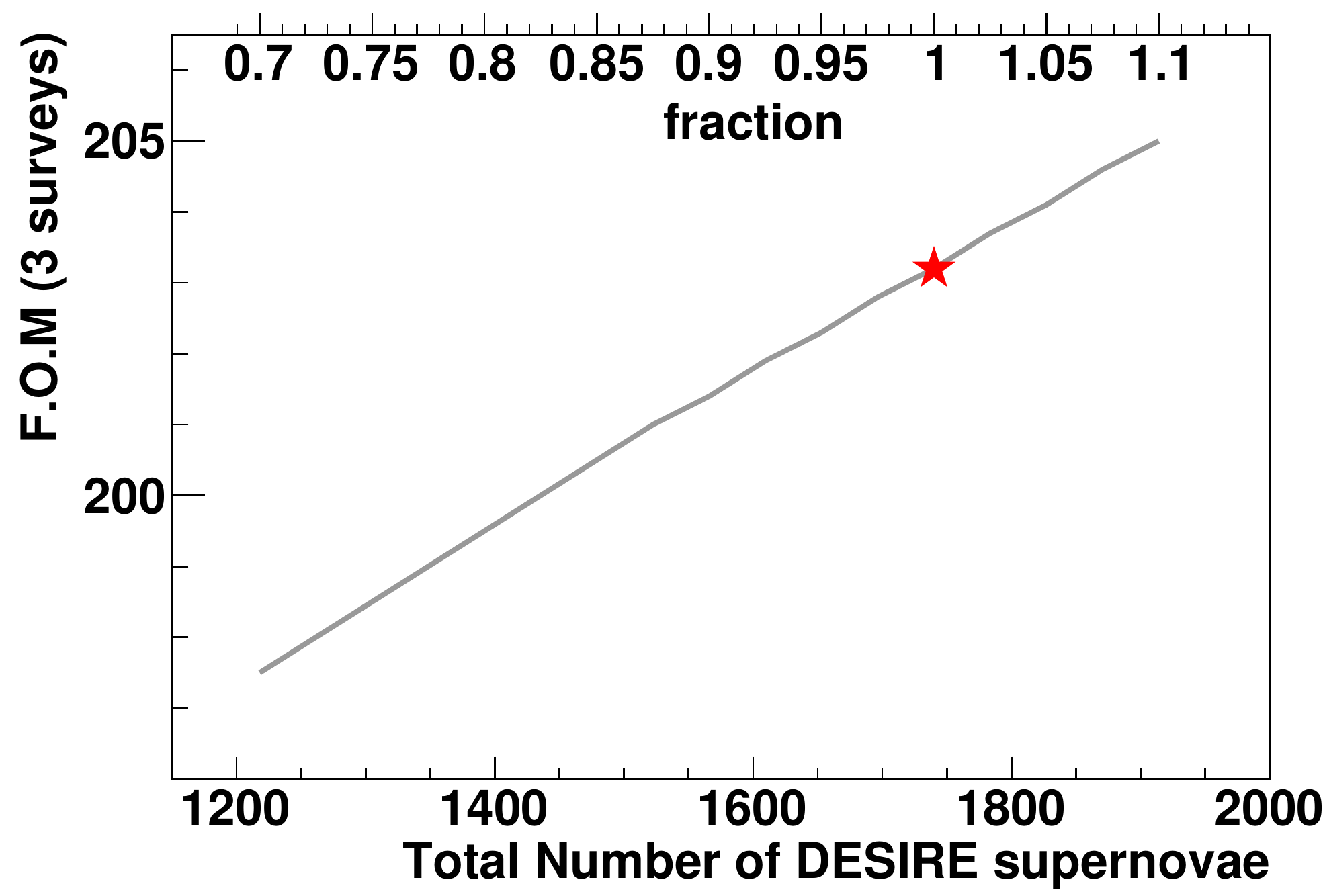}
\caption{FoM for the 3 surveys as a function of the SN statistics in DESIRE. 
The upper horizontal scale is the fraction of events actually entering into the Hubble diagram, with respect to our baseline assumptions. Event rate
measurements at $z>1$ (see \S \ref{sec:sn-simulator}) suggest
higher statistics  by $\sim$ 20 \%) than we assumed, and the efficiency
at getting host redshifts could eliminate 25\% of the events. In any case,
we see that the cosmological performance does not depend critically
on these numbers.
\label{fig:plot_nsn_fom}
}
\end{figure}

The event statistics of the DESIRE survey is primarily 
limited by the amount of time available on Euclid, and is hence
not extensible. It is then important to assess the impact of lower statistics
on the cosmological performance. We remind here that rates at $z>1$ are uncertain
(but we have adopted a conservative approach), and that we have evaluated 
that a massively parallel spectroscopic campaign to collect DESIRE host redshifts
could reasonably target a $\sim$75~\% completion rate (see \S \ref{sec:redshifts}).
We show in figure \ref{fig:plot_nsn_fom} that the cosmological performance
is not severely affected by a significant decrease of the DESIRE event statistics
actually entering into the Hubble diagram.

\subsection{Altering the baseline survey and systematic hypotheses\label{sec:alteration}}

\begin{figure}[h]
\centering
\includegraphics[width=\linewidth]{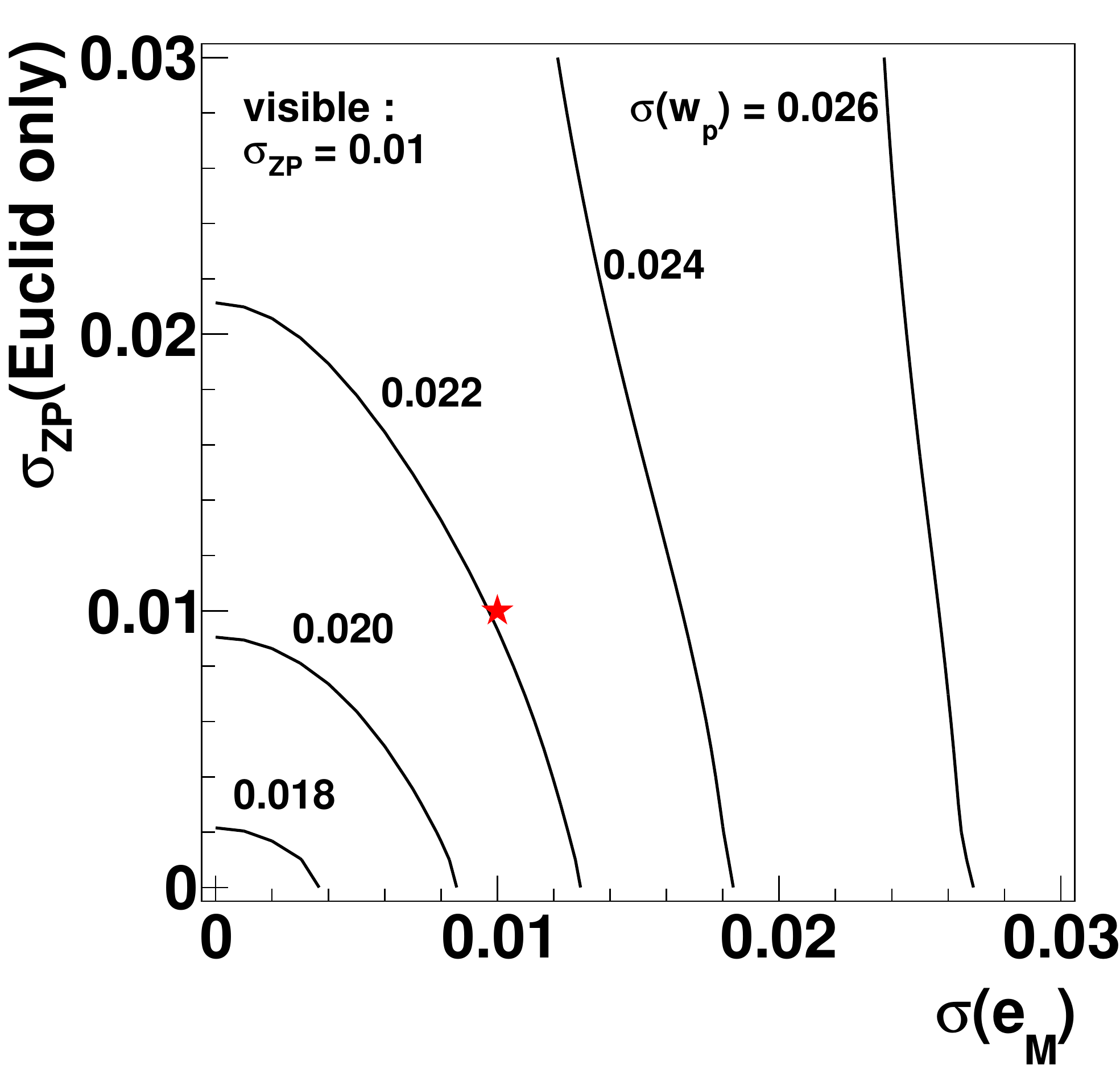}
\includegraphics[width=\linewidth]{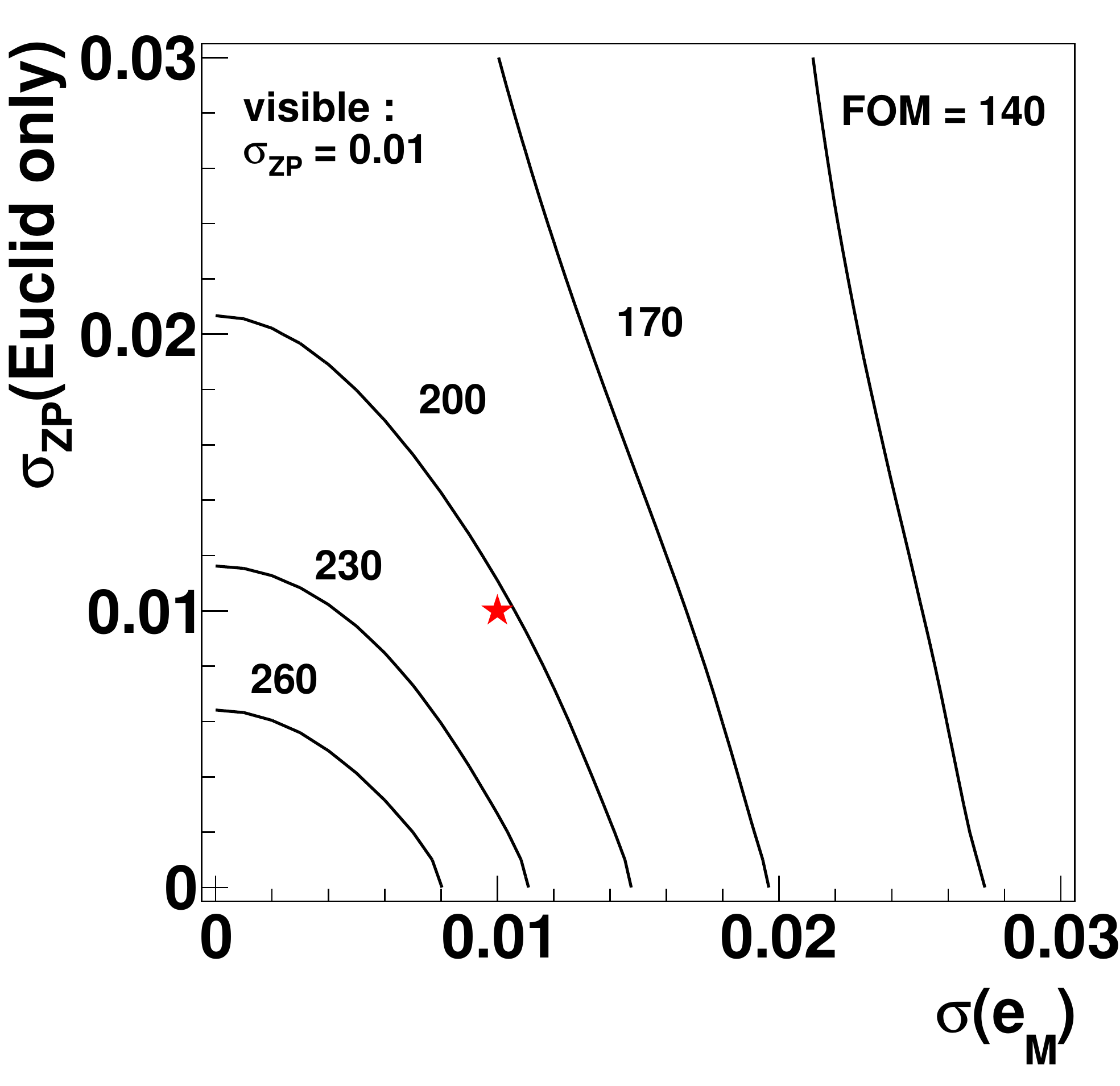}
\caption{Contour levels of $\sigma(w_p)$ (top) and the FoM (bottom) 
as a function of Euclid calibration 
accuracy $\sigma_{ZP}$ (equal for all Euclid filters), and the
distance evolution uncertainty $\sigma(e_M)$ (defined in Eq.~\ref{eq:M_depend_z}). 
The stars indicate our baseline (0.01, 0.01). One can note that significantly worse hypotheses 
do not dramatically degrade the 
capabilities of the proposed surveys.
\label{fig:contours-euclid-only}
}
\end{figure}

The photometric calibration uncertainty (i.e. the zero point
uncertainty) and the evolution uncertainty (equation
\ref{eq:M_depend_z}) constitute the two main performance drivers with a
fixed SN sample size. Fig.~\ref{fig:contours-euclid-only} shows 
the cosmological performance as a function of the size of these systematic
uncertainties. Regarding the photometric calibration, we have varied only
the NIR calibration accuracy (i.e. Euclid's photometric calibration), 
since photometric calibration accuracy in the visible is already better than
what we assumed (see \citealt{Betoule13}).
We note that the performance is reasonably
robust to significant changes in these two uncertainties.

We investigate how the performance varies with statistics in 
Table~\ref{tab:alterations}: the FoM varies roughly as the square root of
the total number of events (rather than linearly without systematics nor external priors). By altering the overall statistics of 
each of our three surveys separately, we show that all three 
contribute similarly to the cosmological precision (as already indicated 
in Table~\ref{tab:cosmo-fom}). The DESIRE part shows the smallest relative change, mostly because
it has the smallest number of events in a first place.
We note that modest improvements
of the SN modelling quality (intrinsic scatter and
colour smearing) significantly improve the 
overall performance. 

\cite{Scolnic13} propose to describe the scatter around the
  brighter-bluer relation using $ \sigma_c = 0.04$ and
  $\sigma_{int}=0$, where we use by default respectively 0.025 and
  0.12; transferring the scatter from brightness to colour also increases $\beta$ from
  3 to $\sim$4. With this extreme setup, we find a FoM of 204,
  i.e. unchanged with respect to our baseline.

\begin{table}[ht]
\caption{Effect of altering some survey parameters.\label{tab:alterations}}
\begin{center}
\begin{tabular}{l|cc|}

{\bf Alteration} &  {\boldmath $\sigma(w_p)$} & {\bf FoM} \\  
\hline
{\bf none} & 0.022 & 203 \\
statistics (all surveys) $\times$ 1.25 & 0.021 & 231 \\
statistics (all surveys) $\times$ 0.75 & 0.024 & 172 \\
low-z $\times$ 1.25 & 0.022 & 218 \\
low-z $\times$ 0.75 & 0.023 & 187 \\
LSST-DDF $\times$ 1.25 & 0.022 & 212 \\
LSST-DDF $\times$ 0.75 & 0.023 & 193 \\
DESIRE $\times$ 1.25 & 0.022 & 208 \\
DESIRE $\times$ 0.75 & 0.022 & 199 \\
$ \sigma_c = 0.015 $ & 0.021 & 223 \\
$ \sigma_{int}=0.10$ & 0.021 & 231 \\
$ \sigma_c = 0.04$ $\sigma_{int}=0$ & 0.022 & 204 \\
\hline
\end{tabular}
\end{center}
\tablefoot{$\sigma_c$ refers to colour smearing (0.025 by default),
$\sigma_{int}$ refers to the Hubble diagram scatter (without any
  shot noise nor colour smearing), set to $0.12$ in our baseline.}
\end{table}

Rather than altering globally the statistics of the three
  proposed surveys, one may study how a small event sample at a given
  redshift improves the cosmological performance as a function of this
  redshift.  With our setup, $z<0.1$ is the most
  efficient redshift range, because we have less than 200 events at
  $0.05<z<0.1$ (see Fig. \ref{fig:nsn}). Adding 200 supernovae at
  $z=0.05$ improves the figure of merit by more than 30. However,
  incorporating a low-redshift sample into the analysis requires 
  that it is measured in three bands in the B,V,R spectral region,
  that the photometry is precisely cross-calibrated with respect to other 
  samples and that this nearby sample is essentially unbiased. The latter probably
  implies to collect it in the ``rolling search'' mode, which is a
  demanding requirement given the sky area that has to be patrolled
  for collecting 200 low-redshift SNe~Ia. We have not incorporated such
  a sample in our forecast, but one could argue that existing facilities
  (e.g. PTF \citealt{PTFPresentation} ; Skymapper \citealt{SkyMapperPresentation}) could deliver it soon. 

\section{Comparison with the DES and WFIRST SN survey proposals}
\label{sec:des-wfirst}

\begin{figure}[h]
\centering
\includegraphics[width=\linewidth]{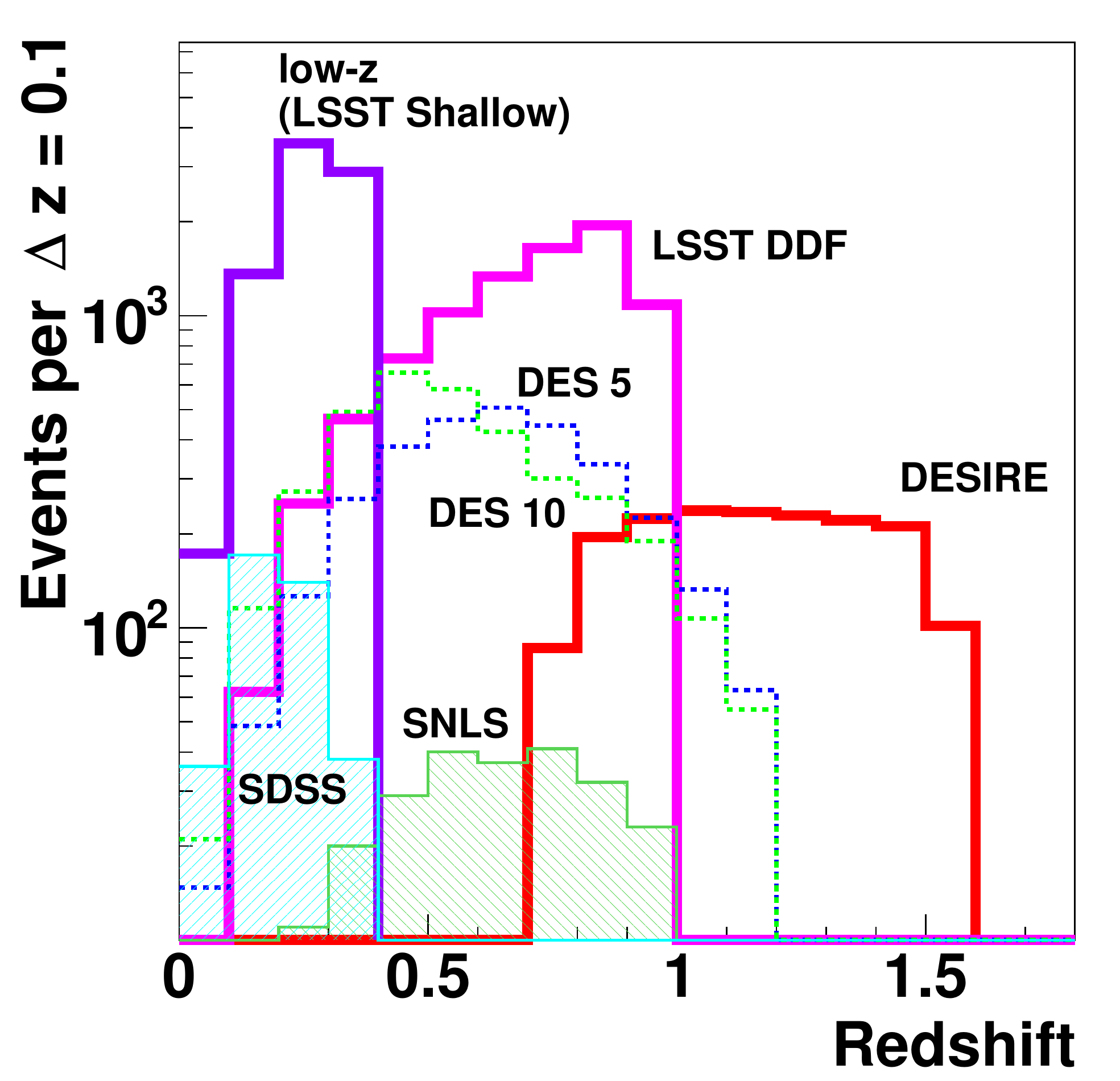}
\caption{Redshift distribution of events for various surveys.
For the SDSS and SNLS, the distributions sketch the total sample
of spectroscopically identified events eventually entering the Hubble diagram.
``DES 5'' and ``DES 10'' refer respectively to the ``hybrid-5'' and ``hybrid-10''
strategies studied in \cite{Bernstein12}, where the baseline is
hybrid-10. ``LSST-SHALLOW'', ``LSST-DDF'' and ``DESIRE'' refer to the three
prongs studied in this proposal.
\label{fig:nsn}
}
\end{figure}

\cite{Bernstein12} present the forecasts for a SN survey to be conducted
within the Dark Energy Survey (DES). This work anticipates about 3000
events with acceptable distances at $0.3<z<1.2$, complemented by a 300-event
nearby sample and 500 events from the SDSS. The presented cosmological
constraints incorporate a ``DETF stage-II
prior''\footnote{We are grateful to R. Biswas for providing us with
  the weight matrix of this prior.}, which accounts for more than
just {\it Planck} constraints: on its own, this prior delivers a FoM of 58.
The forecast does not account for uncertainties arising from SN model
training. In order to compare our findings with this work, we
compute our FoM in the same conditions: 
we temporarily adopt the same external prior, we ignore SN model training uncertainties,
and we let the curvature float. Our assumptions about calibration
uncertainties are already the same as those from \cite{Bernstein12}. 
We find a FoM of 468 for our 3 surveys, to
be compared to 124 for the SN DES survey \citep[][Table 15]{Bernstein12}.
It is hence clear that our proposal constitutes a significant step forward
after DES. We compare the redshift distributions of the DES 
planned observations
with the current samples and our proposal in figure \ref{fig:nsn}.

WFIRST is a NASA project of a NIR wide-field imaging and spectroscopy
mission in space \citep[W12 hereafter]{WFIRSTReport}. The mission is
presented in two versions DRM1 and DRM2 with mirrors of 1.3 and 1.1 m
diameter and durations of 5 and 3 years respectively. In both
instances, the primary mirror is un-obstructed, which not only
enhances its collecting power, but also allows for a more compact PSF
than a conventional on-axis setup with the secondary mirror and its supporting
structure in the beam. The baseline supernova survey (assuming DRM1,
see W12 p. 34) makes use of 6 months of observing time spread over 1.8~y, 
and devotes more than two thirds of its observing time to
low-resolution prism spectroscopy; the remainder is used for
imaging in J,H and K bands every 5 days. This is a dual-cone survey, 
where the deep part targets $0.8<z<1.65$ and
covers 1.8 deg$^2$, and the shallow part targets $z<0.8$ over an area
of 6.5~deg$^2$, to which one should add the contribution 
from the deep survey footprint. The integrations at each visit
last 1500~s and 300~s in the two surveys. The forecast adds a nearby survey
of 800 events (at typically $z<0.1$). 
We thus have three redshift regimes which roughly gather 
(in increasing order) 800, 1400 and 700 events. The high-z part gathers
about half of the statistics targeted by DESIRE, but extends
to higher redshifts. The low-redshift part is very different 
from the one we have sketched: it is first at lower redshift and 
second should be measured  
in significantly redder bands (around 1 $\mu$m) than most current
nearby samples (and our projected low-z part), in order to match the restframe bands of higher redshift
events considered in the project. The intermediate part is also very 
different from our LSST-DDF sketch (or any ground-based $z<1$ SN survey 
in the visible) because it measures in the NIR. 
It is not obvious that the project would significantly benefit
from considering intermediate-redshift events measured in the visible
from the ground, and the forecast concentrates on an
essentially space-based programme.

The anticipated sensitivity of the instrument outperforms Euclid by
more than 0.7 magnitudes: a 1500~s integration with WFIRST reaches
beyond H=26.7 (5 $\sigma$ point source)\footnote{We computed the
  anticipated depth of a 1500~s SN visit from the reported depth H=29.6 of the final
  stack of $\sim$130 epochs (W12 p. 10).} while Euclid remains below 
H=26 (despite its wider H filter). The chosen strategy makes a very efficient
use of this exquisite sensitivity by acquiring low-resolution spectra of all space-based
events, which is not a plausible option for Euclid. 
The quality requirement for light curves is slightly stricter than ours:
S/N$>$15 at maximum light with a 5 day-cadence, while ours translates to
S/N$>$12 at maximum for the same cadence. The WFIRST survey design breaks our
requirement regarding similar restframe wavelengths at all redshifts,
because the span of the measurements in (J,H,K) (about a factor of 2 in wavelength)
is narrower than the redshift range $0.1<z<1.65$ (i.e. $1+z$ varies by about 2.4). Relative distances hence heavily rely on
the SN model and are affected by calibration uncertainties of the training 
sample. In W12, systematic uncertainties of distances to SNe are modelled 
as independent
in $\Delta z = 0.1$ bins with a value that matches the statistical
accuracy from $\sim$50 events in the same bin at low redshift and $\sim$25 
events at high redshift. With these assumptions and {\it Planck} priors, W12 find a FoM
of about 150, which reaches 240 when systematic uncertainties are
halved. The $z>1.5$ part contributes less than 20 to the FoM (Fig.~18
of W12).

\begin{table}[ht]
\caption{Our findings for the WFIRST SN survey performance,
  complemented by 800 nearby events.}
\begin{center}
\begin{tabular}{ccc|cccc|}
\multicolumn{3}{c|}{\bf Assumptions} &  \\
{\bf cal} & {\bf evo} & {\bf train} &  {\boldmath $\sigma(w_a)$} &  {\boldmath $z_p$} &  {\boldmath$\sigma(w_p)$} & {\bf FoM} \\ 
\hline
n & n & n & 0.19 & 0.28 & 0.016 & 323 \\
y & n & n & 0.24 & 0.37 & 0.025 & 170 \\
n & y & n & 0.22 & 0.24 & 0.019 & 242 \\
y & y & n & 0.26 & 0.32 & 0.029 & 132 \\
n & n & y & 0.32 & 0.32 & 0.017 & 180 \\
n & y & y & 0.34 & 0.30 & 0.021 & 137 \\
y & n & y & 1.04 & 0.33 & 0.032 & 30 \\
y & y & y & 1.05 & 0.33 & 0.035 & 28 \\
\hline
\end{tabular}
\end{center}
\tablefoot{``cal'' refers to calibration uncertainties ($\sigma_{ZP}=0.01$).
``evo'' refers to evolution systematics (Eq.~\ref{eq:em_times_z}).
``train'' refers to SN model training from the same sample.
The same quantities for our proposal are shown in Table~\ref{tab:aaa-sys-combinations}.
\label{tab:wfirst}}
\end{table}

In order to compare the SN survey proposed in W12 to the present proposal, we apply
our simulator to the WFIRST SN survey, in particular with our baseline 
systematic uncertainties.  
We have thus performed an approximate 
simulation of the 3-prong survey proposed in W12, and we note that 
we are in a regime where intrinsic fluctuations dominate over shot noise, and
hence the details of the instrument sensitivity are not crucial. With
our assumptions about rates, we find similar overall statistics to W12,
larger for the high-z part by about 200 events, and lower for the
mid-z part by the same amount.  We do not regard this difference as 
important. Table~\ref{tab:wfirst} displays the cosmological
performance we extract from our simulator, which
is again strongly driven by assumptions about systematics at play. When
considering uncertainties induced by photometric calibration and
evolution, we find a FoM of $\sim$ 130, very close to the
value of $\sim$ 150 found in W12, although we have assumed
that uncertainties are correlated across redshifts. Because the 
restframe wavelengths are changing with redshift,
the SN model cannot be extracted from the same sample, as indicated by the dramatic
performance drop in the last rows of Table~\ref{tab:wfirst}.
For the proposed SN survey of W12, the observed SN fluxes as a function of redshift 
can be described by different associations of an SN model (i.e. flux as a
function of wavelength) and a distance-redshift relation, and extracting
both from the data yields large cosmological uncertainties.
These large uncertainties motivate our strategy of extracting distances
from the  same rest-frame region at all redshifts.
The projections in W12 assume, in contrast to ours, that 
the SN model has been developed elsewhere, and that the assumed systematic
uncertainties make provision for all SN model uncertainties.
For our proposal, we get a FoM=266 if we ignore SN model uncertainties
(see Table~\ref{tab:aaa-sys-combinations}).

Recently, a new WFIRST concept has been proposed, relying on
  an existing 2.4m space-quality on-axis primary mirror. A scientific
  programme and an instrument suite taking advantage of this powerful
  telescope have been proposed \citep{WFIRST2.4-final-report}. The
  supernova programme still uses 6 months of observing time but follows
  a different route: SNe are discovered using the wide-field imager
  and their distance are estimated from a series of photometric R$\sim$~100
  spectra ($0.6<\lambda<2\ \mu m$) obtained using an Integral
  Field Unit. The SN programme acquire $\sim$ 7 spectra of the SN at a
  5-day cadence, and the equivalent of 4 epochs for the reference. The
  SN spectra deliver a S/N for synthetic broad-band
  photometry of about 15 per filter at each visit, except for one spectrum at
  maximum light that reaches S/N$\simeq$ 50. Events are selected for
  spectro-photometric follow-up so that the redshift distribution is
  flat at $0.6<z<1.7$ with 136 events per $\Delta z=0.1$, and more
  populated at lower redshifts. The forecast anticipates similar
  contributions of systematics and statistics, using the optimistic
  hypothesis for systematics from W12: the IFU instrument is assumed
  to be easier to calibrate than an imager, and using spectroscopy
  allows one to get rid of cross-redshift K-corrections. The forecast
  does not provide a figure of merit. It seems a priori 
very difficult to complement the proposed
analysis using samples measured in the visible from the ground
through imaging (as in our sketch), both because of the different
measurement technique but also because of the different restframe
wavelength coverage.
  
Both SN proposals for WFIRST aim at covering a redshift range
  wider than the spectral coverage of the instrument, and hence have
  to measure supernovae at different redshifts in different restframe
  spectral ranges. This makes both of them vulnerable to
  inaccuracies of the SN model used to relate these different restframe spectral
  regions. To get around this limitation, one either has to show that
  the incured uncertainties are negligible (our Table~\ref{tab:wfirst}
  indicates that it is not the case), or narrow the redshift range of
  the space project to match the wavelength coverage of the instrument. 
  In this second hypothesis, all considered NIR space missions will have
  to complement their high-redshift samples with lower redshift events
  presumably from wide-field ground-based facilities.

  Regarding the 2.4 m supernova survey project, it still has to be
  demonstrated that measuring flux ratios from an IFU can reach the 
  required accuracy (typically a few $10^{-3}$).  On the other hand, one cannot
  question that a 2.4~m wide-field space mission has a farther reach
  than Euclid for distances to SNe, should it eventually rely on the
  ``traditional'' and established imaging methods. The time line of
  this project remains uncertain.

\section{Astrophysical issues}
\label{sec:astro-issues}
\subsection{Host galaxy stellar mass}
It has been shown that even after applying the brighter-slower and brighter-bluer
relations, residuals of the Hubble diagram are correlated
with host galaxy stellar mass \citep{Kelly10,Sullivan10,Lampeitl10}. Obviously, the host galaxy stellar
mass is a proxy for some physical source of the effect 
yet to be uncovered (see e.g. \citealt{Gupta11,Childress13}),
possibly metallicity \citep{Dandrea11,Hayden13,Pan13}.
In the first analyses considering this effect, 
ignoring it caused a sizable bias on $w$ (e.g. about
0.08 \citealt{Sullivan11}), mostly because low-redshift 
searches favour massive hosts, while rolling searches discover
events regardless of the host properties. The 
`` JLA'' SN sample \citep{Betoule14} is composed at more than 80\%
by the SDSS and SNLS rolling searches, and when
fitted together with  {\it Planck}, ignoring the host mass dependent
brightness shifts $w$ by less than 0.01.
This does not conflict with
the 5 $\sigma$ detection of the host mass dependent brightness on this sample,
but rather indicates that its host mass distribution evolves 
slowly with redshift. In the PS1 SN analysis \citep{Scolnic14}, the host-mass effect turns out to be barely detected.

   The required host stellar mass precision is modest because 
the correction varies slowly with stellar mass (e.g. Fig.~3 of \citealt{Sullivan10}). 
This host stellar mass is estimated 
using galaxy population synthesis models fitted to 
broad-band photometric measurements of host galaxies.
As in past SN surveys, the surveys
we are discussing in this paper offer the opportunity to gather this 
photometric data in typically 5 bands. 
Although
it is likely that our understanding of the phenomenon
will have improved by the time Euclid flies, the data required
to account for the effect by current methods is indeed 
a by-product of the SN surveys, as experienced by current projects. The photometric depth 
obtained by stacking all DESIRE images (\S \ref{sec:ultra-deep-field})
seems sufficient for this purpose, considering that, as done currently,
one can just assign a low stellar mass to apparently host-less SN events.

   As discussed above and in A11, if the understanding
of the effect requires separate models for SN subclasses, and/or separate 
$\alpha$, $\beta$, and ${\mathcal M}$ for different stellar mass
hosts, or some other quantity, as suggested in \cite{Hayden13},
the degradation of cosmological performance is negligible.

If obtaining host redshifts significantly selects among the
  event population (in particular in the DESIRE part), the analysis
  should take care at restoring similar host populations at all
  redshifts, typically in order to ensure that applying or not the
  chosen host correction does not have a serious effect on
  cosmological conclusions. This might in turn reduce the statistics
  of nearby and mid-redshift samples. For these samples, one should
  regard the event statistics we have considered as what is actually
  used for cosmology. We note that for both of these samples, our
  hypotheses are well below what the LSST instrument can plausibly
  deliver within its planned programme.

\subsection{Spectroscopy of ``live'' supernovae and metallicity diagnostics}
\label{sec:spectro-UV}

Almost all SN cosmology works so far have acquired a live spectrum
of their events, but this is impractical for the sample
size we are considering here. We however still 
envisage collecting a sizable sample of
SN+host spectra. This is known to be feasible at $z<1$
\citep[e.g.][]{Zheng08,Balland09,Blondin12}. In the next decade, we
can seriously consider extending the spectroscopic comparisons of SNe
across redshifts and host types to higher redshifts than currently
available: both the JWST and ground-based extremely large telescopes
\citep{Hook12} will provide relatively easy access to mid-resolution
spectroscopy of faint targets (m $\sim$ 26) in the NIR, not practical
with current instruments. These facilities will allow us not only to extend the
spectroscopic comparisons of SNe~Ia to $z=1.5$ and above, but also 
to characterise the contamination of the Hubble diagram across redshifts.

Among spectroscopic diagnostics of the chemical composition of the ejecta, 
assessing the details of the UV flux around 300~nm restframe
is particularly useful to estimate the metallicity of the progenitor 
\citep{Lentz00,Foley13}. UV spectroscopic measurements already exist at 
low redshift \citep{Maguire12}, and at $z\simeq 0.6$ \citep[e.g.][and references therein]{Walker12}. Extending the redshift range of such measurements
should become possible. In A11 (\S 5.2) we proposed photometric 
measurements of SN metallicity through the broadband flux 
at $250 \lesssim \lambda \lesssim 320$~nm restframe. Such measurements 
allow one to control offsets of the distance modulus at the 0.01 level.
In the surveys we are sketching, the data for such measurements
is available at $z\simeq 0.7$ ($g$ band) and again at $z \simeq 1.6$ ($i$ band). 

\subsection{Colour relations and dust extinction}
Although the fact that brightness and colours of supernovae are
related is not debated, the physical source of this relation is still
unclear. There are clear signatures of dust extinction
in spectra of highly reddened events (e.g. \citealt{Blondin09} 
and \citealt{Wang09} and references therein), and some indications that part of the
brighter-bluer relation could be intrinsic to supernovae
\citep[e.g.][]{Foley11}. An intriguing observation is that the
colour distributions seem similar across environments
\citep{Sullivan10, Lampeitl10}, although one would expect less
extinction in passive galaxies than in active ones. \cite{Smith12}
even provide some indication that supernovae in passive hosts are
slightly redder than in star-forming hosts. It is thus likely that
the observed brighter-bluer relation is a mixture of dust extinction
and intrinsic SN variability.  Most of the supernova cosmology analyses eliminate
heavily reddened events, likely to be extincted by dust, because they
are rare and faint, and could be atypical.

Since the brighter-bluer relation is linear in colour vs
magnitude space, and different colours are related by linear
relations \citep[e.g.][]{LeibundgutPhd,ConleySifto08} it is natural
to adopt the formalism of extinction. In our analysis, we have adopted 
an agnostic approach, namely deriving the
``extinction law'' from data, without assuming that it belongs to
the classical forms determined for dust in the Milky Way
\citep{Cardelli89}. In our approach, this law is separated in two
parts: a polynomial function of wavelength, and the $\beta$ parameter of
equation \ref{eq:distance-estimator}. A determination of this law from
spectroscopic SN data has shown that it is a smooth function of
wavelength \citep{Chotard11}, and we use a polynomial function with 10
coefficients to model it. As mentioned earlier, if the extinction law and/or
the $\beta$ parameter have to be determined separately for different
event classes, the impact on the cosmological precision is either very small
(A11) or even null in some cases (A11 Appendix C).

The case for a $\beta$ parameter evolving with redshift is unclear
(see e.g. the discussions in \citealt{Kessler09} and \citealt{Conley11}).
As in A11, we follow an agnostic route and evaluate the extra cost of fitting
different $\beta$ values in redshift bins: we find that fitting separate
$\alpha$ and $\beta$ parameters in $\Delta z=0.1$ bins decreases the FoM
by less than 1.

\section{Other science with DESIRE}
\label{sec:deep-vis-ir}

While the primary motivation for the DESIRE survey is precision
cosmology with distant SNe~Ia, the resulting images will enable a
wealth of other science, both using the time series of images and
using the final deep stacked images, from the bands aimed at
measuring distances, but also from the sharp Euclid visible images
which can be acquired simultaneously. It is beyond the scope of the
present paper to explore in detail all the possible scientific legacy
of DESIRE, but we will just mention a few examples.

\subsection{Transient astrophysics}
The large statistics of distant SNe~Ia can be used to measure their rate
evolution as a function of redshift. When compared to the cosmic star
formation history (SFH), the rate evolution with redshift sets strong constraints on
the Delay Time Distribution (DTD) of SNe~Ia, and therefore
provides information on their progenitors (e.g. \citealt{Perrett12, 
MaozMannucci12,Maoz13,Graur13}).  This analysis will benefit from the comparison with the
large transient statistics now available for the local Universe that
is the harvest of a number of very successful SN searches, e.g. the
Palomar Transient Factory \citep{Rau09} or the Catalina Real-Time
Transient Survey \citep{CatalinaSurvey12}.

Additional constraints on the progenitors scenario (\citealt{Maoz13}) can be obtained by
comparing the SNe~Ia rates with the properties of the parent galaxies as
obtained from broad-band photometry
(e.g. \citealt{Mannucci05,Mannucci06,Sullivan06,Li11}) or spectroscopy
\citep{Maoz12}. Besides the astrophysical interest, this analysis is
important for the cosmological use of SNe~Ia because it can help to
control the systematics related to a possible evolution of these
standard candles.

 As well as SNe~Ia, the DESIRE survey will discover $>$1000
  core-collapse SNe that can also be used as cosmological distance
  indicators. In particular, \cite{HamuyPinto02} found a tight
  correlation between the expansion velocity and plateau magnitude for
  II-Plateau SNe (IIP), which has since been extended to
  cosmologically useful redshifts \citep{Nugent06}. Although
  fainter than SNe Ia, their progenitors are well understood and there
  is excellent potential for IIP to be used as complementary probes of
  the cosmological parameters in the NIR \citep{Maguire10}.

In addition, the statistics of core collapse events can be used as an
independent probe of the cosmic SFH (e.g. \citealt{Dahlen12}) or, if this is known from other
estimators, constrain the stellar initial mass function along with the
mass range for core collapse SN progenitor (e.g. \citealt{Botticella08}). There have been claims of a
mismatch between the current estimate of the SFH and the observed rate
of core collapse SNe that needs to be investigated further \citep{Horiuchi11}.
A proposed explanation is that a large fraction of core collapse SNe
remains hidden in particular in the very dusty nuclear regions of starburst galaxies
\citep{Mannucci07} and correcting for these (e.g. \citealt{Mattila12})
can lead to core-collapse SN rates consistent with the expectations
from the cosmic SFH \citep{Dahlen12}.

In this respect a NIR SN search with Euclid is attractive because of
the reduced effect of dust extinction that will allow us to derive a more
complete census of all types of SNe (e.g. \citealt{Maiolino02,Mannucci03}). It has been shown that the bias in observed rates
due to dust extinction is expected to increase with redshift even for
SNe~Ia (e.g. \citealt{Mannucci07,Mattila12} and references therein) and can be a dominant factor above
$z\sim 1$.

While the most heavily extinguished SNe will remain out of
reach even for Euclid, the DESIRE survey will allow the detection of a
large population of intermediate extinction SNe 
which current optical searches mostly miss. 
All together, DESIRE will significantly
increase the number of core collapse SNe in the highest redshift bins
which are not well sampled now.

Finally, we stress that for the purpose of collecting transient
statistics, parallel observations with the Euclid optical channel (VIS) would
be very valuable. With the combination of IR filters and optical
(unfiltered) monitoring, DESIRE will provide detections, as well as 
light and colour curves that can be used for the transient photometric
classification and therefore extended ground based follow-up is not
required for this purpose.

Aside from SNe, the DESIRE project will provide a unique database for
the study of AGN variability. This can be used to identify AGN, in
particular those of fainter magnitude that are more difficult to
detect with other methods, and hence probe the evolution of the faint
end of the AGN luminosity function up to high redshifts (e.g. \citealt{Sarajedini11}).  
At the same time the data will contribute to our
understanding the physics of AGN variability in the IR spectral
window.  It has also been proposed that the detailed AGN light curves
may allow reverberation mapping measurements of AGN/QSO/SuperEddington
accreting massive Balck Holes. Those are currently being studied as possible
"standard candles" that could extend to larger redshifts the range
provided by SNe~Ia for measuring distances (e.g. \citealt{Kaspi05,Watson11,
Bentz13,WangDu13,Marziani14}.

\subsection{The DESIRE ultra deep field\label{sec:ultra-deep-field}}

By the end of the DESIRE survey, an area of 10 deg$^2$ will have been
imaged 90 times, giving a final stacked depth of 28 to 28.5 mag (AB, 5
sigma point source limit) in i,z,y,J and H bands.

Such an ’Ultra-deep field’ would make a unique legacy, being about
2 magnitudes deeper than the Euclid Deep survey (40 deg$^2$ reaching AB=26) while
JWST will reach deeper limits but on a much smaller area (its survey
capability is constrained by NIRCAM’s FoV of 2.2$\times$4.4 arcminutes$^2$,
simultaneously observed in two bands). Examples of uses for such data
include very high redshift (z$>$8) galaxy and QSO surveys going
approximately two magnitudes fainter down the luminosity function than
the baseline Euclid Deep surveys described in \cite{EuclidRB}. We also
note that the spectroscopic SNe~Ia host sample that would be obtained as
part of DESIRE will provide calibration of deep photometric redshifts within this
ultra-deep Euclid field, which will be of lasting legacy value.

Although Euclid VIS data is not used in the SN light curve analysis
(because of its broad wavelength range), the high spatial resolution
imaging of VIS would be a powerful complement. If the VIS imager were
allowed to integrate while the NIR DESIRE images are collected, the
resulting stacked VIS image would reach R$\sim$28 
(5 sigma for an extended
source of 0.3\arcsec\ FWHM). 
Such deep and sharp Euclid
VIS images would allow deep morphological studies of galaxies in the
field (for which matching depth multicolour data would also be available, see
above). 
Using the several hundred dithered exposures of the field, the resulting
stack can afford a finer pixelisation than the instrument 0.1\arcsec/pixel scale.
Such a deep stack would enable measurement of the location of transients
within their hosts, providing information on possible progenitor
scenarios. In the case of SNe~Ia, the position within the host has been
found to correlate with photometric and spectroscopic properties of
the SNe themselves (\citealt{Wang97}, \citealt{Wang13}), which may
yield further improvements on cosmological parameter constraints.

\section{Summary}
\label{sec:summary}

We have simulated a high-statistics SN~Ia Hubble diagram which
consists of three surveys, which in turn cover the whole redshift
range from $z\sim0$ to $z=1.55$. The high-redshift part relies on deep
NIR imaging from Euclid and concurrent observations in $i$ and $z$
bands from the ground, which consist of monitoring the same 10
${\mathrm deg}^2$ footprint for two seasons of 6 months each. During
each season, Euclid observes the fields approximately half of the time,
so the total survey time on Euclid amounts to 6 months (including $\sim$40\% overheads). 

We have placed sufficiently stringent observing quality requirements
so that all surveys are effectively redshift-limited. We have assumed that
the systematic calibration uncertainties are 0.01 mag (i.e. about
two times larger than current achievements), we have included a correlated irreducible
distance modulus uncertainty to account for possible evolution systematics of
the SN population with redshift, and accounted for both statistical and
systematic uncertainties of the SN model used to fit the light curves.
Despite these conservative assumptions,
we find that this large scale Hubble diagram, when combined with a
1-D {\it Planck} geometrical prior, can deliver stringent purely geometrical
dark energy constraints: a static equation of state is constrained
to $\sigma(w) = 0.022$. We find that the anticipated performance is
fairly robust to changing the assumptions on the size of the leading systematic
uncertainties. DESIRE is therefore an exciting prospect for cosmology,
providing significant constraints on dark energy that are independent of 
Euclid's other probes, while the resulting ultra-deep NIR imaging
would enable a wealth of Legacy science.

\begin{acknowledgements}
We are grateful to the anonymous referee for suggesting subtantial improvements
to the original manuscript. 
  EC, SS and MT acknowledge the grants ASI n.I/023/12/0 “Attivitá
  relative alla fase B2/C per la missione Euclid” and MIUR PRIN 2010-
  2011 “The dark Universe and the cosmic evolution of baryons: from
  current surveys to Euclid”.
\end{acknowledgements}

\appendix
\section{Simulating point source photometry uncertainties}
\label{sec:point-source-photometry}
In order to simulate the precision of SN observations, 
we have to derive the flux measurement
uncertainty from the value of the source flux, the instrument characteristics,
and the observing conditions.
For a point source (SNe are point sources) the expected content of a pixel $p_i$ reads:
\begin{equation*}
p_i = f \psi_i+ (d + s)T
\end{equation*} 
where $f$ is the object flux, $\psi_i$ the PSF at pixel $i$ (i.e.
the fraction of the object flux in this pixel), d is the dark current
per pixel, s is the sky background per pixel per unit time, and T is the exposure time. 
The flux of a supernova is obtained by integrating the (redshifted) SN 
spectrum in the bandpass of the instrument, accounting for the distance.
Expressing all quantities in
electrons, the variance reads:
\begin{equation*}
V_i =  f \psi_i+ (d + s)T + r^2
\end{equation*} 
where r is the rms read noise and the other terms are just Poisson variance.
A least-squares fit of $f$ to the image should minimise:
\begin{equation*}
\chi^2 = \sum_i  \left[ I_i - p_i \right ]^2/ V_i
\end{equation*} 
where $I_i$ are the measured pixel flux values. 
The flux estimator reads:
\begin{equation*}
\hat{f} = \frac {\sum_i  I_i  \psi_i /V_i} {\sum_i \psi_i^2/V_i}
\end{equation*} 
and its variance :
\begin{equation*}
Var(\hat{f}) = \frac {1} {\sum_i \psi_i^2/V_i} 
\end{equation*} 
This flux variance is statistically optimal and it is the expression we use in our simulator. How $V_i$
depends on the object flux determines how $Var(\hat{f})$ behaves at the bright and faint ends. 
At large flux, $Var(\hat{f}) \rightarrow f$, as
expected from Poisson statistics.  Faint sources are those for which
sky and dark current dominate the variance. In this regime, the pixel
variance becomes stationnary ($v_i \equiv v = V_i = (d + s)T + r^2$), 
and the flux variance reads:
\begin{equation*}
Var(\hat{f}) = v \frac {1} {\sum_i \psi_i^2d} 
\end{equation*} 
The rightmost factor has the dimension of an area (expressed in
number of pixels) and is often called the noise equivalent area (NEA).
It summarises the PSF quality for photometry of point sources:
\begin{equation}
NEA = \frac {1} {\sum_i \psi_i^2} \label{eq:NEA-definition}
\end{equation}
with $\sum_i \psi_i = 1$.
This expression accounts for
pixelisation, and is always larger than 1 pixel. For a well sampled
Gaussian, the noise equivalent area reads $4\pi\sigma^2$, where
$\sigma$ is expressed in pixels. The NEA values for the Euclid bands are provided (in arcsec$^2$) in Table~\ref{tab:sky-zp}.  
The Euclid NIR imager is sufficiently
coarsely sampled for the NEA to vary with the source position within a
pixel by 10 to 20 \% r.m.s. Our simulator makes use of a single
position that delivers a NEA representative of the average.

One might note that the above arguments ignore that, in PSF
photometry, one usually has to fit for both position and flux. However,
for an even PSF function, position and flux are uncorrelated, and
fitting the position together with the flux does not degrade the flux variance.

The algebra above applies to measurements in a single image, while
supernova photometry requires to subtract an image of the field
without the supernova (deemed ``reference image''). The impact of this
subtraction is discussed in the next appendix.

\section{Evaluating the influence of a finite reference image depth}
\label{app:ref-depth}

The fluxes of a supernova are obtained by subtracting supernova-free
images from images sampling the light curve. Since the same
supernova-free image (or image set) is subtracted from all SN epochs,
the inferred supernovae fluxes are statistically correlated by the
noise on these supernovae-free images, deemed reference. The SN fluxes
$f^{SN}_i$ (where $i$ indexes epochs) can be written as:
\begin{equation*} 
f^{SN}_i = f_i-f_{ref}
\end{equation*} 
where $f_i$ is the flux in image $i$ and $f_{ref}$ is the flux
measured (at the same position) on the reference image. The covariance
matrix of SN fluxes reads
\begin{equation*} 
C_{ij} \equiv Cov(f^{SN}_i,f^{SN}_j) = \delta_{ij}Var(f_i)+\sigma^2_{ref}
\end{equation*} 
with $\sigma^2_{ref} \equiv Var(f_{ref})$. In matrix notation:
\begin{equation*} 
W \equiv C^{-1} = (D+ \sigma^2_{ref} \vec{1} \vec{1}^T)^{-1} = D^{-1} -  \sigma^2_{ref} \frac {( D^{-1}\vec{1}) (D^{-1} \vec{1})^T}{1 + \sigma^2_{ref}\vec{1}^T D^{-1} \vec{1}}
\end{equation*} 
where $D_{ij} \equiv \delta_{ij} Var(f_i)$, $\vec{1}$ is just a
vector filled with 1's, and we have made use of the Woodbury matrix
identity.  Modern differential photometry techniques
\citep[e.g.][]{Holtzman08} do not explicitly subtract a reference
image, but instead fit a model to images both with and without the
supernova, but this does not change the structure of the output 
SN flux covariance matrix.

The measurements $f_i$ are used to fit the light curve parameters $\theta$
by minimising
\begin{equation*} 
\chi^2 = [A\theta-F]^T W [A\theta-F],
\end{equation*} 
where $F\equiv (f^{SN}_1,\ldots,f^{SN}_n)$ and $A\theta = E[F]$ (E[X] denotes the expectation value of the random variable X). The inverse covariance matrix
of estimated parameters reads
\begin{equation} 
C_{\hat \theta}^{-1} = A^TWA = \sum_i w_i h_i h_i^T - \sigma^2_{ref} \frac{\sum_i w_i h_i \sum_i w_i h_i^T}{1+\sigma^2_{ref} \sum_i w_i}, \label{eq:finite_ref_depth}
\end{equation} 
where $h_i$ are the rows of A (i.e. $h_i = \partial E[f^{SN}_i]/\partial \theta$), $w_i^{-1} \equiv Var(f^{SN}_i)$, and sums run over the considered measurements. 
We use this expression to account for finite reference depth. 
For an amplitude parameter $a$ in a single band (i.e. all light curve points
scale with $a$), we have $h_i(a) \propto  E[f_i] \equiv \phi_i$, so that
the variance of ${\hat a}$, with other parameters fixed, reads
\begin{equation*} 
\left [ \frac{Var({\hat a})}{a^2} \right]^{-1} = \sum_i w_i \phi_i^2 - \sigma^2_{ref}\frac { \left[\sum_i w_i \phi_i \right]^2}{1+\sigma^2_{ref} \sum_i w_i},
\end{equation*} 
where the second term is the contribution of the finite reference depth.

\section{Supernovae in the Euclid deep field(s)}
\label{sec:SN-in-deep}

In this section we study the reach of repeated Euclid standard visits
regarding distances to SNe~Ia.  In the Euclid wide field survey, NIR
photometry is primarily aimed at securing photometric redshifts of
galaxies. This is turned into the requirement $m_{AB}=24$ point
sources are measured at 5 $\sigma$ in each of the three NIR bands,
fulfilled by the NIR photometry from a Euclid ``standard visit'' in
the wide survey \citep{EuclidRB}. The Euclid deep survey is
constructed from repeated standard visits of the same fields, 40
visits being the baseline, both to increase depth and for calibration
purposes. The latter impose that the observation sequence be exactly
the same as in the wide survey.

Following the adopted way to evaluate depth for the Euclid wide
survey, we simulated three exposures of 79, 81 and 48 s each in the y, J
and H bands respectively as a ``standard visit''. The visits indeed
acquire four of these images, but the envisaged dithers between these four
exposures are such that most of the covered sky area indeed only has three
exposures.

\begin{figure}[ht]
\centering
\includegraphics[width=\linewidth]{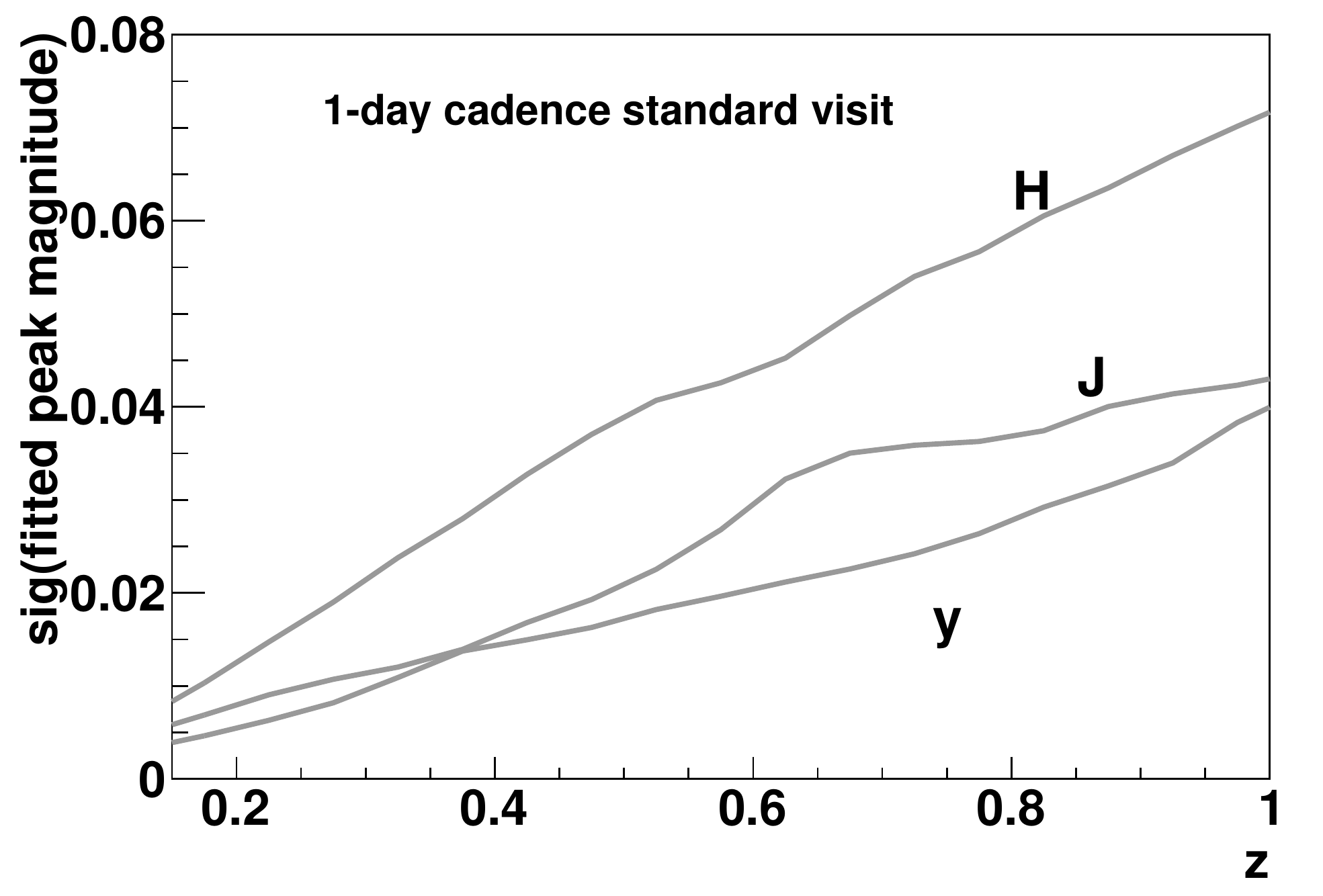}
\caption{Precisions of the fitted amplitude of light curves
with a one-day cadence of standard visits, as a function of redshift.
The requirement of 0.04 is met at $z<\sim 1$ in y and J, 
and $z<0.5$ in H. 
\label{fig:one-day-cadence-amp}}
\end{figure}

We simulated supernovae observed with a one-day cadence which is
probably the fastest possible cadence, and find that the precision of
light curve amplitudes is below 0.04 mag  up to redshifts of
$\sim$1, 0.9 and 0.5 for $y$, $J$ and $H$ respectively, as shown in
Fig.~\ref{fig:one-day-cadence-amp}.  Unfortunately, $H$
being the reddest band, it is most useful at the high end of the
redshift interval, as we aim at covering the same restframe spectral
range at all redshifts. These SN simulations require to model SNe at
wavelengths redder than what SALT2 covers and we have assembled for these
simulations a SN~Ia model in the NIR described in appendix
\ref{sec:saltnir}.

This one-day cadence would allow us to survey about 10 deg$^2$, if
exclusively observing this area, at least for some period of
time. Optimistically assuming that we operate the one-day cadence over
6 months (i.e. visiting the fields 180 times rather than 40 times as
currently envisioned), and integrating events up to $z=1$ (which marginally
meets our quality requirements), we would
collect about 500 SN Ia events, i.e. only about what SNLS collected.

We hence believe that the Euclid deep fields will not deliver
data that allows one to measure a compelling set of SN distances, even 
considering a number of visits far above the Euclid current plans. 

\section{SN Ia model in the rest-frame NIR}
\label{sec:saltnir}
\begin{figure}[ht]
\centering
\includegraphics[angle=90,width=\linewidth]{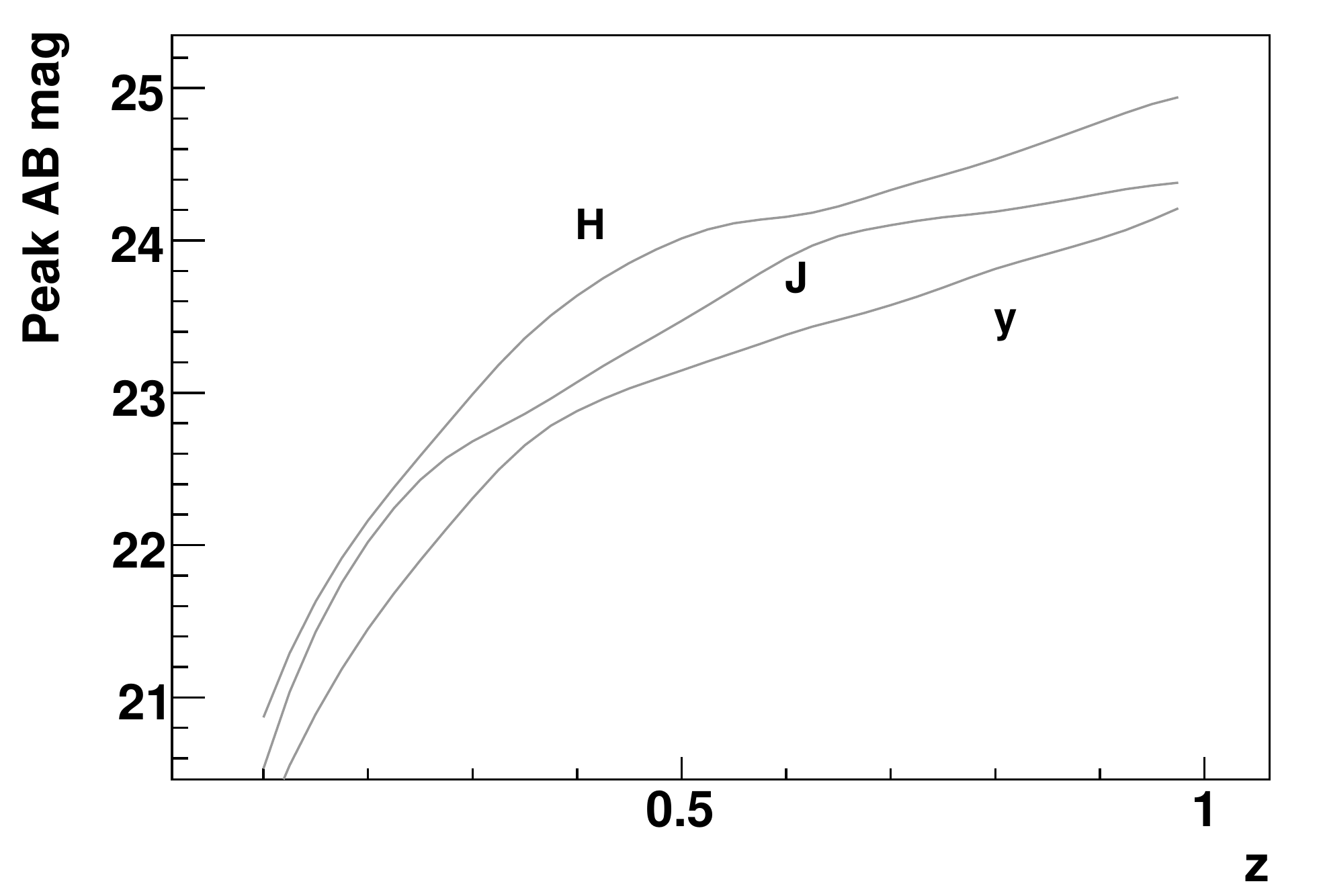}
\caption{Average peak AB magnitude of SNe Ia observed in
the Euclid bands as a function of their redshift, in 
a flat $\Lambda$CDM model, as
predicted by our SaltNIR model, trained on nearby multi-band SN Ia events
from \cite{Contreras10}.
\label{fig:saltnir-peak-mags}}
\end{figure}

Since at low redshift, NIR bands address redder restframe spectral
regions than those covered by SALT2, we developed a simple SN model
designed to deliver realistic amplitudes and light curve shapes up to
almost 2 microns in the rest frame, following the SALT model strategy \cite{Guy05}. It
consists of optimising broadband corrections to an empirical spectral
series in order to reproduce a set of training light curves. Our
``SaltNIR'' model makes use of the E.Hsiao spectral SN template
\citep{Hsiao07}, with broadband corrections derived using the first
release of low-redshift events from the Carnegie Supernova Project
\citep{Contreras10}. The wavelength restframe coverage is
      [330,1800]~nm for the central wavelengths of the simulated
      filters. The training data set misses the restframe z-band which
      hence consists of the spectral template corrected by
      interpolations between i and Y bands. The peak brightnesses
      predicted by this model in Euclid bands are shown in
      Fig.\ref{fig:saltnir-peak-mags} and available in computer-readable form
at \url{http://supernovae.in2p3.fr/~astier/desire-paper/}.

\section{Material for SN-only forecasts\label{sec:binned-matrix}}

We here provide the distance constraints that the proposed observing
sketch could deliver, including all statistical correlations. This can be combined with
other probes as desired. We 
parametrise the distance-redshift relation as linear piecewise
relation parametrised at equidistant pivot points  $z_i = \delta z*i$,
with $i = 0..N$. We define $d_i \equiv H_0 d_M(z_i)/c$ and 
linearly interpolate distance values between the pivot points.
$d_M$ refers to the proper motion distance:
$$
d_M(z) = \frac{c}{H_0 \sqrt{|\Omega_k|}} \mathrm{Sin} \left( \sqrt{|\Omega_k|} \int_0^z \frac{dz'}{H(z')}  \right)
$$
where Sin(x) = sinh(x), x, sin(x) according to the sign of the curvature.
The $d_i$ values define the cosmology (or more precisely the distance-redshift
relation), except for the two first ones,
for which we impose  $d_0 = 0$ and $d_1 = \delta z$. We provide the inverse of the 
covariance matrix of the $d_i$ ($i \geqslant 2$) parameters obtained from
SNe alone, marginalised over all nuisance parameters.

Given an isotropic  cosmological model that defines the proper motion distance
$d_{mod}(z;\theta)$ as a function of some cosmological parameters $\theta$,
we define the residuals to some fiducial cosmology  $\theta_0$ as:
\begin{equation}
R_i = d_{mod}(z_{i+2};\theta)-d_{mod}(z_{i+2};\theta_0)
\end{equation}
where the lowest index of $R$ is $0$. Distances should be understood here
as dimensionless, i.e. $H_0 d_M/c$.
The least-squares constraints expected from SNe around the $\theta_0$ model
simply read:
\begin{equation}
\chi^2 = R^T W R
\end{equation}
where W is the matrix we provide in computer-readable format,
at \url{http://supernovae.in2p3.fr/~astier/desire-paper/}. This SN $\chi^2$ 
can then be added to $\chi^2$   from other probes to obtain overall cosmology constraints.

We have checked that with $\delta z=0.025$, the cosmological
constraints (with our CMB prior) computed 
directly and going through this binned distance scheme 
agree to better than 1\%. With $\delta z=0.025$, there are 63 control points from z=0 to z=1.55,
and we provide a matrix of dimension 61, omitting the first two points as explained above. 
To generate this matrix, we used a flat $\Lambda$CDM model with $\Omega_M = 0.27$, but the current
uncertainties of the distance-redshift relation should not require
to alter this W matrix for cosmologies that yield a realistic 
distance-redshift relation.

\bibliographystyle{aa}
\bibliography{biblio}

\end{document}